\documentclass[12pt,a4paper]{article}
\usepackage{jheppub}
\usepackage{amsmath,amssymb}
\usepackage{float}
\usepackage{comment}
\usepackage{graphicx,xcolor}
\usepackage{euscript}
\usepackage{subfigure,multirow,slashed}
\usepackage{simplewick}
\usepackage{stmaryrd}

\newcommand{\bq}{\begin{eqnarray}}
\newcommand{\eq}{\end{eqnarray}}
\newcommand{\eps}{\varepsilon}

\newcommand{\df}{\mathrm{d}}

\newcommand{\braces}[1]{[\hspace{-0.5mm}[#1]\hspace{-0.5mm}]}

\def \e  {\mathop{\rm e}\nolimits}

\numberwithin{equation}{section} 

\allowdisplaybreaks

\newcounter{MBQ}

\newcounter{YJQ}

\newcounter{XWQ}

\title{\boldmath 
Renormalization of the next-to-leading-power 
$\gamma\gamma \to h $ and $gg\to h$ soft quark functions}

\author[a]{Martin Beneke,} 

\author[a]{Yao Ji}

\author[a]{and Xing Wang}

\subheader{
\begin{flushright}
{\small
TUM-HEP-1501/24\\
March 26, 2024 \\
}
\end{flushright}
}

\affiliation[a]{
   Physik Department T31, James-Franck-Stra\ss e 1, 
   Technische Universit{\"a}t M{\"u}nchen,\\
   D-85748 Garching, Germany}

\abstract{We calculate directly in position space 
the one-loop renormalization kernels of the soft operators 
$O_\gamma$ and $O_g$ that appear in the soft-quark contributions to, respectively,
the subleading-power $\gamma\gamma\to h$ and $gg\to h$ form factors mediated by 
the $b$-quark. We present an IR/rapidity divergence-free definition for 
$O_g$ and demonstrate 
that with a correspondent definition of the 
collinear function, a consistent factorization theorem is recovered. 
Using conformal symmetry techniques, we 
establish a relation between the evolution kernels of the leading-twist heavy-light 
light-ray operator, whose matrix element defines the $B$-meson light-cone distribution amplitude (LCDA), and $O_\gamma$ to all orders in perturbation theory. Application of this relation allows us to bootstrap the kernel of $O_\gamma$ to the two-loop level.  We construct an ansatz for the kernel of $O_g$ at higher orders. We test this ansatz against 
the consistency requirement at two-loop and find they differ only by a particular constant. 
}


\notoc
\begin{document}

\maketitle

\newpage

\tableofcontents
\setcounter{page}{1}
\allowdisplaybreaks


\section{Introduction}
\label{sect:intro}

Soft functions -- vacuum expectation values of soft fields -- appear in 
factorization theorems for QCD high-energy scattering whenever soft effects 
do not simply cancel. In leading power in the high-energy expansion, 
the soft fields appear through semi-infinite light-like Wilson lines 
defined as
\begin{equation}
\label{eq:Wilsonlines}
	Y_{n}(s) = \widehat{\rm P}\exp\left[ig_s\,T^a\int_{-\infty}^{0 }\df 
\lambda\,n\cdot A^a(s+\lambda n)\right],
\end{equation}
where $\widehat{\rm  P}$ is the path-ordering, $g_s$ denotes the strong coupling, 
and $T^a$ the $SU(N_c)$ generator in the representation of the energetic particle 
interacting with the soft gluon. This is a consequence of the fact that at leading 
power soft gluons couple only through the eikonal vertex $i g_s T^a n^\mu$, 
where $n^\mu$ is the direction of the four-momentum of the particle. A well-known example is the soft operator 
\begin{equation}
\label{eq:DYsoftfunction}
\frac{1}{N_c}\,\mbox{Tr} \,
\mathbf{\bar{T}}(Y^\dagger_{n_-}(x^0) Y_{n_+}(x^0)) 
\,\textbf{T}(Y^\dagger_{n+}(0) Y_{n_-}(0))
\end{equation}
for the Drell-Yan process $q\bar{q} \to \gamma^*$ near the 
kinematic threshold \cite{Korchemsky:1993uz}. 
$\mathbf{{T}}$ denotes the time-ordering. A similar soft operator 
weighted by the final-state measurement function is relevant to the two-jet limit of 
event shapes in $e^+ e^-$ annihilation.

In soft-collinear effective theory (SCET), the soft Wilson lines arise from the field redefinition that removes the eikonal interaction from the leading-power soft-collinear Lagrangian \cite{Bauer:2000yr}. This transformation does not remove the soft-gluon interactions from the sub-leading 
power Lagrangian \cite{Beneke:2002ph,Beneke:2002ni}, resulting in soft 
field-strength insertions $Y^\dagger_n F_{\mu\nu} Y_n$ in the soft 
operators. Generalized soft-operators of this type appeared first in the 
context of heavy-quark physics \cite{Beneke:2004in,Lee:2004ja,Bosch:2004cb} 
and in the context of high-energy scattering in 
\cite{Moult:2018jjd,Beneke:2018gvs}. In addition, there exist 
sub-leading power soft operators that contain soft light quarks, 
rather than gluon field-strength insertions. These soft functions 
are key ingredients in factorization formulas at next-to-leading power (NLP), 
but relatively little is known about them beyond their definition 
and leading-order expressions from direct computation. An exception 
is the Drell-Yan threshold for which both the soft-gluon and soft-quark 
NLP soft function have been computed at $\mathcal{O}(\alpha_s^2)$ 
\cite{Beneke:2019oqx,Broggio:2021fnr,Broggio:2023pbu}. Their 
anomalous dimension is however only known at the level of 
the leading double pole from direct computation, and only for 
the case of the soft gluon field 
insertion \cite{Beneke:2018gvs,Beneke:2019mua}. Developing methods 
to compute the anomalous dimension (renormalization-group kernels) 
of generalized soft functions is a prerequisite for extending 
resummation at NLP to the next-to-leading logarithmic accuracy.

In this work, we focus on the NLP soft quark operators $O_\gamma$ and 
$O_g$ that appear in the factorization formula for the 
subleading-power $\gamma\gamma\to h$ and $gg\to h$ form factors mediated by 
the bottom quark \cite{Liu:2019oav,Liu:2020wbn,Liu:2022ajh}. 
In the context of the factorization formula, the 
leading-order renormalization group (RG) kernels for both operators 
have been inferred from RG consistency \cite{Liu:2020eqe,Liu:2022ajh}, 
that is, from the requirement that the full process is factorization 
scale independent and the prior knowledge of the RG kernels and 
anomalous dimensions of all other functions in the factorization 
formula. For the case of  $O_\gamma$, the consistency has been 
confirmed by a direct calculation of the kernel \cite{Bodwin:2021cpx}, 
but a corresponding result for the non-abelian generalization $O_g$ is 
not available. In the following we show that the RG kernels for 
both soft operators can be obtained with position-space 
renormalization methods by a relatively straightforward computation. 
The case of $O_\gamma$ relevant to $\gamma\gamma\to h$ turns out 
to be very similar to the renormalization of the leading-twist heavy-light 
quark light-ray operator \cite{Lange:2003ff,Braun:2003wx} that defines the 
so-called $B$-meson LCDA. The $gg\to h$ soft operator $O_g$ presents 
a complication, related to ``colour charges at infinity''.  In this 
respect, it resembles the QED-generalization of the $B$-meson LCDA 
for decays into two electrically charged light 
particles  \cite{Beneke:2019slt,Beneke:2022msp}.

Recent years have witnessed rapid developments in 
the application of conformal symmetry to perturbative QCD 
calculations (see~\cite{Braun:2003rp} for an early pedagogical
introduction and the recent treatment~\cite{Braun:2013tva} with technical details in~\cite{Braun:2016qlg}).
The basic idea is to consider QCD at the critical point in  
non-integer dimensions where the $\beta$ function vanishes 
identically, allowing one to take advantage of the conformal symmetry to the fullest extent. 
In this work, we extend the use of conformal symmetry, for the first time, to the study of non-string operators $O_\gamma$ and $O_g$, where the quark fields are located on different light rays.

The outline of this paper is as follows: We set up the definitions of the soft operators in Sec.~\ref{sec:setup}, then calculate 
the RG kernel for $O_\gamma$ in Sec.~\ref{sect:gamma}. The result 
coincides with the one from \cite{Bodwin:2021cpx}, but our derivation 
is considerably simpler. The renormalization of the soft operator  $O_g$ 
for the soft-quark contribution to the $gg\to h$ form factor is 
presented in Sec.~\ref{sect:gluon}. The naive definition of the  
operator leads to a rapidity-divergent anomalous dimension, which 
requires an extra subtraction. As a consequence, the factorization 
formula presented in \cite{Liu:2022ajh} needs a rearrangement 
between its soft and (anti-) collinear parts that resembles the factorization of electromagnetic effects in two-body $B$-meson 
decays to electrically charged mesons \cite{Beneke:2019slt,Beneke:2020vnb}. Regarding the renormalization, the soft-quark operator for 
$\gamma\gamma\to h$ is essentially the ``double-copy'' of the 
leading-twist $B$-LCDA operator. 
We exploit this observation in  Sec.~\ref{sec:CS} 
to infer its two-loop RG kernel from the known result for the 
$B$-LCDA \cite{Braun:2019wyx} using the conformal symmetry technique. We construct an ansatz for the kernel of $O_g$ at higher orders and test it against 
the consistency requirement at two-loop. We conclude in Sec.~\ref{sect:conclusions}. 
Three appendices provide further technical details. In particular, 
in App.~\ref{app:jet} we demonstrate that the collinear matching coefficients, 
which enter the subleading-power factorization formulas, remain unaltered and gauge-independent in the presence of an off-shell infrared regulator in general 
covariant $R_\xi$ gauge.

\section{Setup}
\label{sec:setup}

The soft-quark functions appearing in the Higgs production through a light-quark loop in two photon ($\gamma\gamma\to h$) and two gluon ($gg\to h$) fusion at the subleading power are defined as vacuum matrix elements of operators made of semi-infinite, light-like Wilson lines 
\eqref{eq:Wilsonlines} and soft-quark fields. The bare operator definitions in position 
space read,
\begin{eqnarray}
\label{def:Oga}
&& O_\gamma(s, t) 
= {\bf T}\,\Big\{\bar{q}(tn_-)Y_{n_-}(t)Y^\dagger_{n_-}(0) \frac{\slashed{n}_-\slashed{n}_+}{4}Y_{n_+}(0)Y^\dagger_{n_+}(s)q(sn_+)\Big\} \,,\\[0.1cm]
\label{def:Oguns}
&& O_g^{\rm uns}(s, t) =
{\bf T}\,\Big\{\bar{q}(tn_-)Y_{n_-}(t)T^aY^\dagger_{n_-}(0) \frac{\slashed{n}_-\slashed{n}_+}{4}Y_{n_+}(0)T^bY^\dagger_{n_+}(s)q(sn_+)\Big\} \,,\quad
\end{eqnarray}
for $\gamma\gamma\to h$ and $gg\to h$, respectively, 
where $n_\pm^\mu$ are two light-like vectors, $n_\pm^2=0$, 
normalized to $n_-\cdot n_+=2$. $q$ denotes the light quark field. 
The superscript ``uns" on $O_g(s,t)$ indicates that it is an \textit{unsubtracted} operator corresponding to the definition in~\cite{Liu:2022ajh}\footnote{Unlike \cite{Liu:2022ajh}, we define $O_g^{\rm uns}$ without an overall factor of 
$g_s^2$. Neither does $O_\gamma$ include an overall factor $e_b^2$. }.  
A necessary modification will be explained in Sec.~\ref{sect:gluon}. Notice that the two adjoint colour indices $a,b$ in ${O}^{\rm uns}_g$ are kept open, and we keep them implicit for convenience. Here, the light-like Wilson lines $Y_{n_\pm}$ given in~\eqref{eq:Wilsonlines} are in the fundamental representation of $SU(N_c)$. 
While the operator $O_\gamma$ stays the same for the $h\to\gamma\gamma$ decay process due to \eqref{eq:Ogammadef} below, the Wilson lines $Y_{n_\pm}$ in $O^{\rm uns}_g$ for $h\to gg$ decay extend from $0$ to $+\infty n_\pm$ instead, as required for the 
coupling of soft gluons to out-going (anti-) collinear 
gluon fields. The anomalous dimensions, however, remain the same under the exchange of initial and final state due to the time-reversal symmetry of QCD. 

In the $\gamma\gamma\to h$ case, the semi-infinite Wilson lines combine to form finite-length segments as in the case of light-cone string operators, 
\begin{equation}
\label{eq:Ogammadef}
O_\gamma(s, t) = {\bf T}\,\Big\{ \bar{q}(tn_-)[tn_-, 0]\frac{\slashed{n}_-\slashed{n}_+}{4}[0, sn_+]q(sn_+)\Big\}\,,
\end{equation}
with
\begin{equation}
[z_1n_\pm, z_2n_\pm] \equiv \widehat {\rm P} \exp\left[ig_s T^a \int^{z_1}_{z_2} \df\lambda \, n_\pm \cdot A^a(\lambda n_\pm)\right]\, .
\end{equation}

The case of $O_g^{\rm uns}(s, t) $ is fundamentally different due to two inserted colour matrices corresponding to colour charges at the infinity, which arise from the 
energetic back-to-back gluons in the initial state of the $gg\to h$ process. We 
can bring  $O_g^{\rm uns}(s, t)$ into a form that resembles $O_\gamma(s,t)$ by 
applying the identity
\begin{equation}
Y_{n_{\pm}}(x) T^{a} Y_{n_{\pm}}^{\dagger}(x) = \left(\mathcal{Y}_{n_{\pm}}(x)\right)^{ab} T^{b}
\end{equation}
to convert the fundamental Wilson lines to adjoint ones, defined as
\begin{equation}
\label{eq:adjW}
\left(\mathcal{Y}_{n_{\pm}}(x)\right)^{ab} = \widehat{\rm P}\exp\left[-g_s\,f^{abc}\int_{-\infty}^{0 }\df \lambda\,\e^{\lambda (- i\delta_\pm+0^+)}n_\pm\cdot A^c(x+\lambda n_\pm)\right]\,,
\end{equation}
where $0^+$ guarantees convergence of the integral and is related to the pole prescription for the in-going eikonal propagator. We further introduced the infrared (rapidity) regulator $\e^{-i\lambda \delta_\pm}$ with $\delta_i\in \mathbb{R}$ in the definition of the semi-infinite adjoint Wilson lines~\cite{Echevarria:2015byo} . 
In momentum space, this regulator is equivalent to modifying the eikonal 
propagator generated by the $\lambda$-integration as 
\begin{equation}
\label{eq:eikp}
\frac{i}{n_\pm\cdot \ell+i0^+}\mapsto \frac{i}{n_\pm\cdot \ell + \delta_\pm +i0^+}
\end{equation}
with $\ell$ denoting the soft momentum absorbed by the Wilson line. 
The need for this regulator will become clear in Sec.~\ref{sec:Og}. In the following, we often drop the $+i0^+$ with the implicit understanding that $\delta_\pm$ means $\delta_\pm + i0^+$. 
The $\delta_\pm$ regulator can be further related to the off-shellness of the energetic external gluons from which the semi-infinite Wilson lines are induced. More specifically, the eikonal propagators describing soft gluon absorption in the (anti-) collinear sector are constructed as follows:
\begin{align}
    &\frac{i n_+\cdot p_{c}}{(p_{c}+\ell)^2+i0^+} \to \frac{i}{n_-\cdot \ell +\frac{p_{c}^2}{n_+ \cdot p_{c}} +i 0^+}
    \equiv \frac{i}{n_-\cdot \ell+\delta_- + i0^+}\,,\notag\\
    &\frac{i n_-\cdot p_{\bar c}}{(p_{\bar c}+\ell)^2+i0^+} \to \frac{i}{n_+\cdot \ell +\frac{p_{\bar c}^2}{n_- \cdot p_{\bar c}} +i 0^+}
    \equiv \frac{i}{n_+\cdot \ell+\delta_+ + i0^+}\,.
\end{align}
Hence we identify the relations
\begin{equation}
\label{eq:delta-offshell}
   \delta_+ = \frac{p_{c}^2}{n_- \cdot p_{c}} + i0^+ ,
   \qquad \delta_- = \frac{p_{\bar c}^2}{n_+ \cdot p_{\bar c}} + i0^+ ,\,
\end{equation} 
where $p_c$ ($p_{\bar c}$) denotes the (anti-) collinear external momentum. These relations will play an important role in Sec.~\ref{sec:fac}. 
It is then possible to rewrite $O_g^{\rm uns}(s,t)$ as 
\begin{eqnarray}
O_g^{\rm uns}(s, t) &=& 
		{\bf T}\,\Big\{\bar{q}(tn_-)[tn_-, 0]\left(\mathcal{Y}_{n_-}(0)\right)^{ac} T^{c}\frac{\slashed{n}_-\slashed{n}_+}{4}\left(\mathcal{Y}_{n_+}(0)\right)^{bd} T^{d}[0, sn_+]q(sn_+)\Big\}
\nonumber\\
&& \hspace*{-2.2cm}={\bf T}\,\Big\{
		\bar{q}(tn_-)\left(\mathcal{Y}_{n_-}(t n_-)\right)^{ac} T^{c}[tn_-, 0]\frac{\slashed{n}_-\slashed{n}_+}{4}[0, sn_+]\left(\mathcal{Y}_{n_+}(sn_+)\right)^{bd} T^{d}q(sn_+)\Big\}
.\qquad
	\label{eq:Ogdef}		
\end{eqnarray}
The expression in the second line can be derived directly from the subleading power SCET Lagrangian insertion without applying the soft-decoupling transformation. 
In practice, we find it is more convenient to adopt this expression for calculating the renormalization factor of the operator $O_g^{\rm uns}(s,t)$.

In phenomenological studies of Higgs decay or production through a bottom-quark loop, the soft functions enter the factorization formulae as the vacuum matrix elements of the relevant operators in momentum space. The required Fourier transform reads, \footnote{Compared to~\cite{Liu:2019oav,Liu:2022ajh}, apart from the overall $e_b^2$ for $S_\gamma$ and $g_s^2$ for $S_g$, we neglect an overall prefactor $(4\pi)^{-\eps} e^{\eps\gamma_E}/(\pi i)$ for brevity, which will not affect renormalization.}
\begin{equation}
\label{eq:softdef}
	S_{\gamma,g}(w) = \frac{1}{2\pi i}\,{\rm Disc}_{\ell_-\ell_+}\int_{-\infty}^\infty \df s\,\int^{\infty}_{-\infty}\df t\, \e^{ is\ell_+-it\ell_-}\,\langle 0 |  O_{\gamma,g} (s,t) | 0 \rangle\,.
	\end{equation}
	The fact that the momentum-space soft functions depend on the single variable $w=\ell_-\ell_+$ is a consequence of reparameterization/Lorentz invariance. 
``${\rm Disc}$" denotes the discontinuity of the Fourier transform, and $\ell_\pm\equiv n_\pm\cdot\ell>0$ are the soft quark light-cone momentum components conjugate to $s$ and $t$, respectively. 
They are positive since their physical interpretation corresponds to twice the soft quark energy as the multipole expansion puts the soft quark position on the light cone.  We anticipated that the Fourier transform for the \textit{subtracted} soft gluon operator $O_g(s,t)$ must be used instead of the \textit{unsubtracted} one,  $O_g^{\rm uns}(s,t)$. 
The definition for $O_g(s,t)$ will be given later in~\eqref{def:Osb}.

To convert the anomalous dimension from position space to momentum space, it proves to be more convenient to perform the Fourier transform at the operator level using the expression~\cite{Braun:2017liq}
\begin{align}\label{eq:Fourier}
    O_{\gamma,g}(w) &= \int_{-\infty}^\infty \df s\,\int^{\infty}_{-\infty}\df t\, \e^{ i(s-i0^+)\ell_+-i(t+i0^+)\ell_-}\, O_{\gamma,g} (s,t)\, .
\end{align}
Similar to the case of the $B$-meson LCDA, it is necessary to take $s$ ($t$) as a variable in the lower (upper) half of the complex plane, hence the replacement $s\to s-i0^+$ ($t\to t+i0^+$) in the Fourier integral. This arises directly from $\ell_\pm>0$ dictating the inverse transformation to be\footnote{In some cases \cite{Beneke:2020vnb,Beneke:2022msp,Huang:2023jdu} renormalization can extend the light-cone momentum fractions to the negative domain. This, however, does not occur for the present case as the positivity of $\ell_+$ and $\ell_-$ is demanded by the factorization formulae~\cite{Liu:2022ajh}.}
\begin{align}\label{eq:Fourier2}
O_{\gamma,g}(s,t) = \int^\infty_0\frac{\df\ell_+}{2\pi}\int^\infty_0  \frac{\df\ell_-}{2\pi} \e^{-is\ell_++it\ell_-}\, O_{\gamma,g} (w)\, ,
\end{align}
which technically constitutes a Laplace transform. Then, the $\mp i0^+$  prescription for $s$ and $t$ in~\eqref{eq:Fourier} follows from the requirement of the existence of the inverse Laplace transform, which guarantees the invertibility between the soft operators in momentum space $O_{\gamma,g}(w)$ and position space $O_{\gamma,g}(s,t)$. In the following we implicitly assume that $s$ ($t$) means $s-i0^+$ ($t+i0^+$). 

The renormalization of these soft operators corresponds to integrating out the quantum fluctuations above the soft scale $\mu$ (of order $m_b$ in phenomenological applications) on which the renormalized soft operators depend. This can be achieved in the background field method \cite{Abbott:1981ke}, which separates fields into (soft) background (with momenta $p<\mu$) and quantum (with momenta $p>\mu$) components,
\begin{align}\label{eq:sqexp}
    q(x) = q_s(x,\mu) + q_q(x,\mu)\, ,\qquad A^\mu(x) = A_s^\mu(x,\mu) + A_q^\mu(x,\mu)\, .
\end{align}
Integrating out the ``quantum" modes in a gauge-invariant operator ${\cal O}$ leads to an effective operator ${\cal O}_{\rm eff}$ that only depends on the ``soft background" degrees of freedom,
\begin{align}~\label{eq:Pint}
    {\cal O}_{\rm eff}(q_s,A_s) = \int {\cal D}[q_q,\bar{q}_q, A_q]\,{\cal O}(q_s+q_q,A_s+A_q)\e^{iS[q_s+q_q,A_s+A_q]-iS[q_s,A_s]}\, ,
\end{align}
where $S$ is the QCD action. The advantage of the background field method is that it retains the gauge invariance of ${\cal O}_{\rm eff}$ in the evaluation of the path integral~\eqref{eq:Pint}. Due to the presence of the soft background fields, in the simplest case, the  conventional gauge-fixing condition needs to be modified to 
\begin{align}\label{eq:bgg}
    D^{ac}_\mu[A_s]A^{c,\mu}_q(x) = (\partial_\mu \delta^{ac} + g_sf^{abc} A^{b}_{s,\mu})A_q^{c,\mu} =0\, ,
\end{align}
where $D_\mu[A_s]$ is the covariant derivative acting on the quantum component in the presence of soft background. To the lowest order in the soft background field $A_s$, the gauge-fixing condition~\eqref{eq:bgg} coincides with the Feynman gauge. Eq.~\eqref{eq:bgg} serves as a simple demonstration of the background field method principle, but it is important to note that the gauge-fixing condition can be generalized to accommodate the general covariant ($R_\xi$) and axial (e.g., light-cone) gauges as well~\cite{Abbott:1981ke}. This allows us, in practice, to adopt the $R_\xi$ gauge in our calculations for integrating out the quantum fields. The expression for the action $S[q_s+q_q,A_s+A_q]-S[q_s,A_s]$ in the background gauge~\eqref{eq:bgg} and the associated Feynman rules can be found in \cite{Abbott:1981ke}. Since \eqref{eq:Pint} is invariant under local gauge transformations of soft background fields, the background technique allows us to choose gauges for the soft background and quantum fields independently.

\section{Renormalization of \texorpdfstring{$O_\gamma(s, t)$}{}}
\label{sect:gamma}

\begin{figure}[t]
\centering
  \includegraphics[width=14cm]{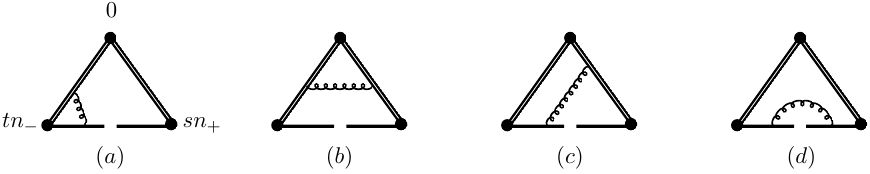}
\caption{\label{fig:Ogamma} One-loop corrections to $O_\gamma(s, t)$. Double lines denote the two finite Wilson line segments, the two horizontal solid lines correspond to the soft quark fields, and the curly lines represent gluon propagators. Mirror diagrams to $(a)$ and $(c)$, the self-energy of the external quarks, and the self-energy of the light-like Wilson lines (which vanishes in $R_\xi$ gauge) are not shown.}
\end{figure}

Since the Wilson lines in $O_\gamma(s, t)$ are finite in length, no rapidity divergences exist. While the potential infrared singularities are naturally regularized by the non-vanishing off-shellness (or mass) of the soft quark, the UV divergence is exclusively captured by dimensional regularization. In this section, we provide details on renormalizing the soft operator directly in the position space, which turns out to be rather compact and easy to generalize to higher loops. The one-loop contributions to $O_\gamma(s, t) $ are depicted in Fig.~\ref{fig:Ogamma}. There are also mirror ones for diagrams $(a)$ and $(c)$. Diagram $(a)$ and its mirror give rise to non-local terms,\footnote{The terminology here refers to the form of the contribution in position space, either simply multiplicative (local), or as a convolution (non-local).} diagram $(b)$ gives rise to a local cusp term, diagram $(c)$ and its mirror contribute in background $R_\xi$ gauge only when $\xi\neq 1$, and diagram $(d)$ does not have UV poles. To simplify the notation, we also use $q(p)$ and $\bar{q}(p)$ to denote the external quark and anti-quark fields in momentum space. We can put either the quark mass or the off-shellness (but not both) to zero from the start since we focus on UV poles, and we choose to drop the quark mass. 

\subsection{Position-space calculation}

Inserting the QCD interaction Lagrangian once, expanding the $n_-$-Wilson line to first order and performing Wick contractions, diagram $(a)$ reads (we drop the subscripts $s$ and $q$ for brevity, and abbreviate $\bar{u}=1-u$),
\begin{equation}
	    \label{eq:IOgammaa}
		\begin{aligned}
		    \contraction[0.8em]{I_a = (ig_s)^2\int\df^Dz\,\bar{q}(z)\,}{A}{^a(z)T^aq(z)\,\bar{q}(tn_-)\int_0^1\df u\,t\,(n_-\cdot)}{A}
		    \contraction[0.6em]{I_a = (ig_s)^2\,\int\df^Dz\bar{q}(z)\slashed{A}^a(z)T^a\,}{q}{(z)\,}{q}
			I_a &= (ig_s)^2\,\int\df^Dz\,\bar{q}(z)\slashed{A}^a(z)T^aq(z)\,\bar{q}(tn_-)\int_0^1\df u\,t\,(n_-\cdot A^b)(utn_-)T^b\frac{\slashed{n}_-\slashed{n}_+}{4}q(sn_+)\\
			&= -g_s^2 C_F \mu^{2\varepsilon}\frac{\e^{\varepsilon\gamma_E}}{(4\pi)^\varepsilon}\int_0^1\!\df u\,t\!\int\frac{\df^Dp}{(2\pi)^D}\!\int\frac{\df^Dl}{(2\pi)^D}\e^{it(\bar{u}n_-\cdot l-un_-\cdot p)}\frac{\bar{q}(p)\,\slashed{n}_-\slashed{l}}{l^2(p+l)^2}\frac{\slashed{n}_-\slashed{n}_+}{4}q(sn_+)\,.
		\end{aligned}
	\end{equation}
We employ dimensional regularization with $D=4-2\varepsilon$ for the 
UV divergences and the  momentum-space $\overline{\rm MS}$ scheme for renormalization, which amounts to subtracting poles in $\varepsilon$.	
The $n_+$-Wilson line $[0,sn_+]$ does not contribute to the integral $I_a$ 
hence is not written for brevity. The Wick contractions are between quantum fields and the integration over $u$ comes from separating the Wilson line into soft background and quantum modes~\cite{Balitsky:1987bk},
\begin{equation}
\label{eq:WLdec}
[tn_-,0] = [tn_-,0]_{s} + i g_s \,t \int^1_0 \df u\, [t n_-, u tn_-]_{s} (n_-\cdot A^b_q)(utn_-)T^b[utn_-,0]_s + \cdots\, .
\end{equation}
The two external quark fields $\bar q(z)$ and $q(sn_+)$ in the first line of~\eqref{eq:IOgammaa} are soft, and we employed the light-cone gauge $n_-\cdot A_s(x)=0$ 
for the soft background fields. This sets the soft background Wilson line in the $n_-^\mu$ direction to $1$ and drastically simplifies the calculation. As explained earlier, we effectively employ the Feynman gauge in detailing our calculations for integrating out the \textit{quantum} fields (loop integrals) and comment on the results in general $R_\xi$ gauge. After some algebra 
we obtain
\begin{align}
\label{eq:IOgammaacont}
I_a &= g_s^2C_F\mu^{2\varepsilon}\frac{\e^{\varepsilon\gamma_E}}{(4\pi)^\varepsilon}\int_0^1\df u\,t\int_0^1\df\alpha \int\frac{\df^Dp}{(2\pi)^D}\int\frac{\df^Dl}{(2\pi)^D}\e^{-i(ut+\alpha t\bar{u})n_-\cdot p+it\bar{u}n_-\cdot l}\notag\\
			&\quad
\phantom{g_s^2C_F\mu^{2\varepsilon}\e^{\varepsilon\gamma_E}}
\times \frac{\alpha\,\bar{q}(p)\slashed{n}_-\slashed{p}}{\left[l^2+\alpha\bar{\alpha} p^2\right]^2}\frac{\slashed{n}_-\slashed{n}_+}{4}q(sn_+)\notag\\
			&= \frac{\alpha_s(\mu)}{4\pi}\frac{2C_F}{\varepsilon}\int_0^1\df u\,\frac{u}{\bar u}\Big[\bar{q}(u tn_-)-\bar{q}(tn_-)\Big]\frac{\slashed{n}_-\slashed{n}_+}{4}q(sn_+) \notag\\
			&= \frac{\alpha_s(\mu)}{4\pi}\frac{2C_F}{\varepsilon}\int_0^1\df u\left[\frac{u}{\bar u}\right]_+ O_\gamma(s, ut) +\mathcal{O}(\varepsilon^0)\, ,
\end{align}
where only $1/\varepsilon$ divergent terms are kept after the second equal sign. The pole arises from the integral over the loop momentum $l$, while integrating over $p$ facilitates Fourier transform back into position space. The non-vanishing off-shellness $p^2$ of the external quark provides the necessary IR regulator. 
In the last step, we elevated the result to operator form without presuming any gauge fixing condition on the soft background fields. This is justified by the gauge invariance of the computation with the background field method. 
Hereafter, we adopt the standard plus-prescription defined as
\begin{align}
	    \int^1_0\df u\,[f(u)]_+ g(u) \equiv \int^1_0\df u\,f(u)\Big(g(u)-g(1)\Big)\, ,
	\end{align}
where $g(u)$ is an arbitrary test function. The contribution from the mirror diagram 
to $(a)$ equals \eqref{eq:IOgammaacont} with $O_\gamma(s, ut)\to 
O_\gamma(us, t)$ on the right-hand side. The result \eqref{eq:IOgammaacont} for diagram  $(a)$ and its mirror diagram remain the same in background $R_\xi$ gauge.
	
Next, we look at the cusp contribution from diagram $(b)$, which is local in position space:
\begin{align}
	\label{eq:IOgammab}
I_b &= -2ig_s^2C_F\mu^{2\varepsilon}\frac{\e^{\varepsilon\gamma_E}}{(4\pi)^\varepsilon}\,\bar{q}(tn_-)\frac{\slashed{n}_-\slashed{n}_+}{4}q(sn_+)\int_0^1\df u\int_0^1\df v\,st\,\int\frac{\df^Dl}{(2\pi)^D}\frac{\e^{-il\cdot(utn_--vsn_+)}}{l^2} \notag \\
		&= - \frac{\alpha_s(\mu)}{4\pi}C_F\,\left[\frac{2}{\varepsilon^2}+\frac{2\ln \left(st\mu^2\e^{2\gamma_E}\right)}{\varepsilon}\right]O_\gamma(s,t)+\mathcal{O}(\varepsilon^0)\,.
	\end{align}
 Hereafter, $\ln(st\mu^2\e^{2\gamma_E}) \equiv \ln(\mu(s-i0^+)\e^{\gamma_E}) + \ln(\mu(t+i0^+)\e^{\gamma_E})$ serves as a shorthand notation without ambiguities regarding the branch cuts of the logarithms. 
Unlike diagram $(a)$, Wilson lines of different directions participate in the interaction, making it impossible to remove the soft background Wilson lines completely by a gauge fixing. Therefore, technically, one has to first expand each Wilson line using~\eqref{eq:sqexp}, then Wick contract between terms with a single quantum field from each direction $n_\pm^\mu$ 
in the expansion using the background-field Feynman rules, 
and finally resum the infinite series. We can, however, bypass all these complications by focusing on computing the corrections to $A_s$ at the lowest order (namely, setting $A_s=0$ in the Wilson lines) knowing that the soft background Wilson lines ensuring gauge invariance will fall in place automatically from higher-order $A_s$ contributions in the  background field technique due to 
its manifest gauge invariance. 
We apply this trick in calculating diagrams hereafter. 
The expression in the first line of~\eqref{eq:IOgammab} comes from Wick-contracting the two quantum gluon fields, $n_-\cdot A_q(utn_-)$ and $n_+\cdot A_q(vsn_+)$ from the Wilson line expansion~\eqref{eq:WLdec} setting $A_s^\mu=0$. The result is then matched to the full gauge-invariant operator $O_\gamma$ yielding the last line of~\eqref{eq:IOgammab}. In general $R_\xi$ gauge, the result for diagram $(b)$ reads,
\begin{align}
    I_b^{(\xi)} &= - \frac{\alpha_s(\mu)}{4\pi}C_F\,\left[\frac{2}{\varepsilon^2}+\frac{2\ln \left(st\mu^2\e^{2\gamma_E}\right)-(1-\xi)}{\varepsilon}\right]O_\gamma(s,t)+\mathcal{O}(\varepsilon^0)\,.
\end{align}

By setting the background gluon field $A_s$ directly to $0$, diagram $(c)$ in Feynman gauge gives rise to the loop integral
\begin{eqnarray}
I_c &=& g_s^2C_F\mu^{2\varepsilon}\frac{\e^{\varepsilon\gamma_E}}{(4\pi)^{\varepsilon}}\,s\int_0^1\df u\int\frac{\df^Dp}{(2\pi)^D}\int\frac{\df^Dl}{(2\pi)^D}\,\e^{i[t n_-\cdot l-usn_+\cdot (p+l)]}
\nonumber\\
&&\hspace*{3cm}\times\,\frac{\bar{q}(p)\,\slashed{n}_+\slashed{l}}{l^2 (p+l)^2 }\frac{\slashed{n}_-\slashed{n}_+}{4}q(sn_+)\notag\\
    &=&-\frac{i\alpha_s(\mu)C_F}{\pi}\mu^{2\varepsilon}\,\e^{\varepsilon\gamma_E}\,s \int\frac{\df^Dp}{(2\pi)^D}\,(n_+\cdot p)\int^1_0\df u\int^1_0 \df \alpha\,\alpha\,\e^{-i(u s\bar\alpha n_++\alpha t n_- )\cdot p}\notag\\
&&\hspace*{0mm}\times\left(-\frac{ust}{\alpha\bar\alpha p^2}\right)^{\varepsilon/2}K_{\varepsilon}\left(2\sqrt{-p^2\alpha\bar\alpha u s t}\right) \bar q(p)\frac{\slashed{n}_-\slashed{n}_+}{4}q(sn_+)\, ,
\label{eq:Iogammac}
\end{eqnarray}
where $K_n(z)$ denotes the Bessel-$K$ function.
While the virtuality $p^2\neq 0$ regulates the IR behaviour of the integral, 
the UV is tamed by the off-light-cone distance between $\bar q(tn_-)$ and $[0,sn_+]$ entering the phase factor. We use the asymptotic behaviour  
\begin{align}
\label{eq:BKas}
    K_\varepsilon(z) \stackrel{z \to 0}\to 2^{-1+\varepsilon}\Gamma(\varepsilon)z^{-\varepsilon}+2^{-1-\varepsilon}\Gamma(-\varepsilon)z^\varepsilon\, ,\qquad\qquad K_0(z) \stackrel{z \to 0}\to -\ln z\, ,
\end{align}
to check that the remaining integrals over $u$ and $\alpha$ do not produce 
singularities from the endpoints $u\to 0$, $\alpha\to 0,1$.  
This argument allows us to conclude that diagram $(c)$ and its mirror do not 
contribute to the renormalization of $O_\gamma(s, t)$ in Feynman gauge. However, diagram $(c)$ and its mirror diagram do contribute in the general $R_\xi$ gauge, which will become important in Sec.~\ref{sect:gluon}. The $R_\xi$ gauge result reads,
\begin{align}\label{eq:Ic}
    I_c^{(\xi)} &= -\frac{\alpha_s(\mu)}{4\pi\varepsilon}  C_F (1-\xi) O_\gamma(s, t)+\mathcal{O}(\eps^0)\, ,
\end{align}
and the mirror diagram contributes identically. 

Finally, the one-loop integral for diagram $(d)$ in Feynman gauge reads,
\begin{eqnarray}
\label{eq:IOgammad}
I_d &=&-ig^2C_F\mu^{2\varepsilon}\frac{\e^{\varepsilon\gamma_E}}{(4\pi)^{\varepsilon}}\int\frac{\df^Dp_1}{(2\pi)^{D}}\int\frac{\df^Dp_2}{(2\pi)^{D}}\int\frac{\df^D l}{(2\pi)^D} \e^{i[t(l+p_1)\cdot n_- - s(l+p_2)\cdot n_+]}
\nonumber\\
&&\times\, \,\bar q(p_1)\gamma_\mu \frac{\slashed{l}+\slashed{p}_1}{(l+p_1)^2} \frac{\slashed{n}_-\slashed{n}_+}{4}\frac{\slashed{l}+\slashed{p}_2}{(l+p_2)^2}\gamma^\mu q(p_2)\,\frac{1}{l^2}\,.
\end{eqnarray}
Similar to diagram $(c)$, the IR-finiteness of diagram $(d)$ is a consequence of the non-vanishing off-shellness $p_1^2,p_2^2\neq0$ of the two external quarks. The superficial (here) logarithmic 
UV divergence of the integrand in the second line is tamed by the off-light-cone phase factor $\e^{i[t n_-\cdot l-usn_+\cdot (p+l)]}$. 
Thus, diagram $(d)$ does not contribute to the 
renormalization of $O_\gamma(s,t)$. In this case, this remains true in general $R_\xi$ gauge. 

In addition, one must account for the $\overline{\rm MS}$ 
quark field renormalization constant. In the general $R_\xi$ gauge, it reads
\begin{equation}
	\label{eq:Zq}
	q^{\rm bare}=Z_q^{1/2}q^{\rm ren} = \left[1-\frac{\alpha_s(\mu)}{4\pi}\frac{C_F}{2\varepsilon}\xi+\mathcal{O}(\alpha_s^2)\right] q^{\rm ren}\, .
\end{equation}
Summing up all contributions in $R_\xi$ gauge, including the mirror diagrams and the quark field renormalization, and then multiplying an overall factor $(-1)$, we find that the $\xi$-dependence cancels, resulting in the one-loop gauge-invariant counter-term/renormalization factor
\begin{eqnarray}
\label{eq:Ogammare}
O_\gamma(s,t;\mu) &=& O^{\rm bare}_\gamma(s,t) + 
\frac{\alpha_s(\mu)C_F}{4\pi} \int_0^{1}  \df u  \left[\left(\frac{1}{\varepsilon^2}+\frac{2\ln\left(st\mu^2\e^{2\gamma_E}\right)+1}{2\varepsilon}\right)\delta(1-u)\right.\notag\\
&& \hspace*{0cm}\left.-\frac{2}{\varepsilon}\left[\frac{u}{\bar u}\right]_+\right]\big(O^{\rm bare}_\gamma(us,t)+O^{\rm bare}_\gamma(s,ut)\big)\,.
\end{eqnarray}
The superscript ``bare'' indicates the bare operator (in terms of renormalized fields). The position-space anomalous dimension / evolution kernel for $O_\gamma(s,t)$ at one-loop is then immediate:
\begin{eqnarray}
\label{eq:1Lgamma-pos}
   [\gamma_\gamma O_\gamma](s,t;\mu) &= &-\frac{\alpha_s C_F}{\pi}\,\Bigg\{
   -\left(\ln\left(st\mu^2\e^{2\gamma_E}\right)+\frac12\right)O_\gamma(s,t;\mu) \\
   &&\hspace*{15mm} +
   \int^1_0\df u\,\left[\frac{u}{\bar u}\right]_+ \big(O_\gamma(us,t;\mu)+O_\gamma(s,ut;\mu)\big)\Bigg\}+{\cal O}(\alpha_s^2)\, .
   \quad\notag
\end{eqnarray}
The operator $O_\gamma$ then satisfies the following renormalization group equation (RGE) in position space,
\begin{equation}
\label{eq:RGE-pos}
\frac{\df }{\df\ln\mu}O_\gamma(s,t, \mu) = -[\gamma_\gamma O_\gamma](s,t;\mu)\, .
\end{equation}

We observe that the one-loop anomalous dimension in position space is rather simple. Most importantly, it essentially constitutes a double-copy of 
the evolution kernel of the leading-twist $B$-meson LCDA \cite{Lange:2003ff,Braun:2003wx}, 
and we will investigate this observation further in Sec.~\ref{sec:GP}. On the one hand, our position space calculational method is straightforwardly applicable to other operators and higher loops. On the other hand, position-space formalism allows us to take advantage of conformal symmetry techniques, which are useful not only for solving the renormalization-group equations (RGEs) but also for performing higher-order calculations. 

\subsection{Anomalous dimension in momentum space}
\label{sec:renoralizationOg}

In phenomenological studies, the momentum-space representation of the renormalization factor/anomalous dimension is often preferred. We therefore apply the Fourier transform to~\eqref{eq:Ogammare}, which can be accomplished straightforwardly by employing 
\begin{eqnarray}
\label{eq:FT1}
    &&\int_{-\infty}^{+\infty}\frac{\df z}{2\pi}\,\e^{i\omega z}\ln(i\mu(z-i0^+)\e^{\gamma_E})\int_0^{+\infty}\df\omega'\,\e^{-i\omega'(z-i0^+)}\tilde{O}(\omega')\notag\\
		&& \hspace{1cm}= \,\int_0^{+\infty}\df\omega'\left[-\delta(\omega'-\omega)\ln\frac{\omega}{\mu}-\omega\left[\frac{\theta(\omega-\omega')}{\omega(\omega-\omega')}\right]_+\right]\tilde{O}(\omega')\, ,\\
&&\int_{-\infty}^{+\infty}\frac{\df z}{2\pi}\,\e^{i\omega z} \int_0^1\df u\frac{u}{\bar{u}}\int_0^{+\infty}\df\omega'\left[\e^{-i\omega'u(z-i0^+)}\tilde{O}(\omega')-\e^{-i\omega'(z-i0^+)}\tilde{O}(\omega')\right]\notag\\
&&\hspace{1cm}= \,\int_0^{+\infty}\df\omega'\left[\delta(\omega'-\omega)+\omega\left[\frac{\theta(\omega'-\omega)}{\omega'(\omega'-\omega)}\right]_+\right]\tilde{O}(\omega')	\label{eq:FT2}\, ,
\end{eqnarray}
where $\tilde O(\omega')$ represents a momentum space ``test function". To apply the above formulae, it is necessary to follow the pole prescription $s\to s-i0^+$, $t\to t+i0^+$ established in~\eqref{eq:Fourier}. For the latter it is helpful to take complex conjugate of~\eqref{eq:FT1} and~\eqref{eq:FT2} to accommodate the $+i0^+$ prescription. 
We adopt the standard plus-prescription in the momentum space defined as,
\begin{align}
\int^{+\infty}_0 \df\omega'\,[g(\omega, \omega')]_+ f(\omega') \equiv \int^{+\infty}_0 \df\omega'\, g(\omega, \omega')  [f(\omega')-f(\omega)]\, ,
\end{align}
where $f(\omega)$ is a test function. 
Following the procedure outlined in appendix~\ref{app:FT} for performing the Fourier transform yields the momentum-space renormalization factor $Z_\gamma$ which reads,
\begin{eqnarray}
\label{eq:Zsp}
Z_{\gamma}(w,w') &=& \delta(w-w')+\frac{\alpha_s(\mu)C_F}{2\pi}\Bigg\{\left[\frac{1}{\varepsilon^2}-\frac{L_w}{\varepsilon}-\frac{3}{2\varepsilon}\right]\delta(w-w')-\frac{2}{\varepsilon}w\Gamma(w,w')\Bigg\}\, ,
\nonumber \\
&&\\[-1cm] 
\nonumber
\end{eqnarray}
where $L_w =\ln\frac{w}{\mu^2}$ and 
\begin{equation}\label{eq:LNK}
    \Gamma(w,w')=\left[\frac{\theta(w-w')}{w(w-w')}+\frac{\theta(w'-w)}{w'(w'-w)}\right]_+
\end{equation}
is the same distribution that appears in the anomalous dimension of the leading-twist $B$-meson LCDA in momentum space~\cite{Lange:2003ff}. 

The anomalous dimension $\gamma_\gamma$ that controls the scale dependence of the vacuum expectation value of the soft operator $O_\gamma$~\eqref{eq:softdef}, $S_\gamma$, through the renormalization-group equation (RGE)
\begin{equation}
	\label{eq:RGE}
	\frac{\df }{\df\ln\mu}S_\gamma(w, \mu) = -\int_0^\infty \df w' \,\gamma_\gamma(w,w') S_\gamma(w', \mu)\, ,
\end{equation}
is then immediate from~\eqref{eq:Zsp},
\begin{align}
	\label{eq:ADp}
	\gamma_\gamma(w, w') &=  -\frac{\alpha_s(\mu)C_F}{\pi}\,\bigg[\Big(L_w+\frac{3}{2}\Big)\delta(w-w')+2 w \Gamma(w,w')\bigg] +\mathcal{O}(\alpha_s^2)\, .
\end{align}
This concludes the calculation of the one-loop renormalization and anomalous dimension of the soft-quark photon operator $O_\gamma$. Eq.~\eqref{eq:ADp} 
coincides with the result previously obtained in \cite{Bodwin:2021cpx}, but has been 
 derived from standard position-space methods without going through transverse-momentum unintegrated functions. 
The position-space formalism proves to be even more advantageous for calculating the anomalous dimension of the more complicated soft-quark gluon operator, as we will detail in the subsequent section.

\section{Renormalization of \texorpdfstring{$O_g(s, t) $}{}}
\label{sect:gluon}

\begin{figure}[t]
  \centering
  \includegraphics[width=14cm]{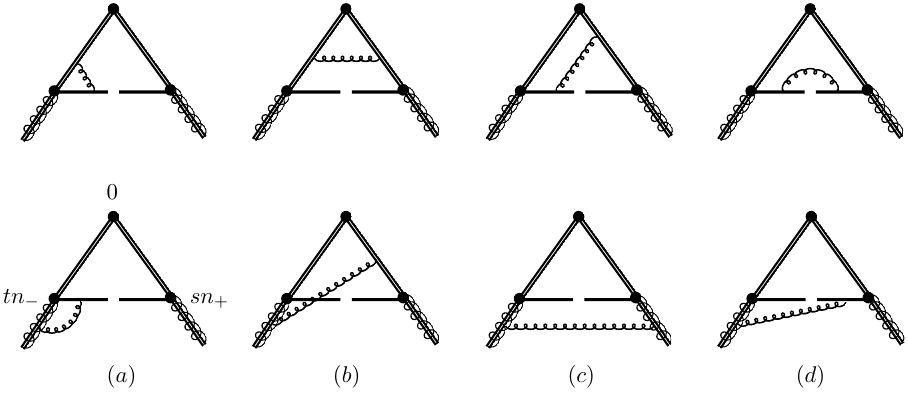}
  \caption{\label{fig:Og}Diagrams responsible for $O_g^{\rm uns}(s, t)$ renormalization at one loop using the representation in the second line of~\eqref{eq:Ogdef}. The double lines superposed with curly ones are semi-infinite adjoint Wilson lines. Mirror diagrams to $(a), (b)$ and $(d)$ are not shown. Contributions from diagrams in the first row are immediately obtainable from eq.~\eqref{eq:Ogammare} with appropriate changes in the colour factor. }
\end{figure}

In this section, we extend the calculation of anomalous dimension to the soft operator 
relevant to the $gg\to h$ case. The one-loop diagrams 
contributing to the renormalization of $O^{\rm uns}_g(s,t)$ 
in the form of~\eqref{eq:Ogdef} are shown in Fig.~\ref{fig:Og}. 
The first row displays diagrams that also appeared for $O_\gamma(s,t)$ 
and their contributions can be obtained from the previous calculation as 
follows: The colour factor $C_F$ in diagram $(a)$ of Fig.~\ref{fig:Ogamma} and its mirror diagram is replaced by $C_F-C_A/2$, while the colour factor $C_F$ in diagram $(b)$ remains the same. 
The colour factor of diagram $(c)$ and its mirror changes to $C_F-C_A/2$, and the contribution of diagram $(d)$ remains zero.
This is a consequence of the colour generators placed between the soft quarks and the fundamental finite-length Wilson lines displayed in the second line of~\eqref{eq:Ogdef}.

There are additional one-loop corrections (shown in the second row of the figure) due to gluon exchanges involving the semi-infinite adjoint Wilson lines  dictated by the second line of~\eqref{eq:Ogdef}.
Due to the two semi-infinite Wilson lines, $O_g^{\rm uns}(s,t)$ exhibits rapidity divergences. Consequently, one needs to separate UV poles from the singular behaviour in rapidity by introducing additional regulators. In this work, we introduce an off-shellness $\delta_\pm$ as in~\eqref{eq:adjW}, \eqref{eq:eikp} to regulate the rapidity divergence induced by the semi-infinite Wilson lines. The presence of the semi-infinite Wilson lines and the consequent introduction of the $\delta_\pm$ regulators can potentially interfere with gauge invariance as will be seen below. Hence it is instructive to employ the general covariant $R_\xi$ background field gauge for the one-loop calculation of $O_g$. 

\subsection{Unsubtracted operator \texorpdfstring{$O_g^{\rm uns}(s,t)$}{}}
\label{sec:Og}

As mentioned earlier, the total one-loop corrections generated by diagrams in the first row of Fig.~\ref{fig:Og} is immediate from Sec.~\ref{sect:gamma}. For the pole part of the sum of all diagrams, including the quark field renormalization  \eqref{eq:Zq}, we find in $R_\xi$ gauge 
\begin{align}
    I_{r_1}^{(\xi)} =&\,\frac{\alpha_s(\mu)}{4\pi}\frac{2}{\varepsilon}\left(C_F-\frac{C_A}{2}\right)\int_0^1\df u \left[\frac{u}{\bar u}\right]_+\left(O_g^{\rm uns}(us,t)+O_g^{\rm uns}(s,ut)\right)\notag\\
	&-\frac{\alpha_s(\mu)}{4\pi}\left\{C_F\left[\frac{2}{\varepsilon^2}+\frac{2\ln\left(st\mu^2\e^{2\gamma_E}\right)}{\varepsilon}\right] -\frac{C_A}{\varepsilon}(1-\xi)+\frac{C_F}{\eps}\right\}O_g^{\rm uns}(s,t)\, ,
\end{align}
which is IR/rapidity finite. Note that the $\xi$-dependence proportional to $C_F$ cancels between diagrams $(b)$, $(c)$ and its mirror\footnote{Keep in mind that we are referring to the first row of Fig.~\ref{fig:Og} but we use labels in Fig.~\ref{fig:Ogamma}.} in Fig.~\ref{fig:Ogamma}, and the quark field renormalization. However the $\xi$ dependence proportional to $C_A$, which originates from diagram $(c)$ in Fig.~\ref{fig:Ogamma} and its mirror diagram, still remains in the sum of the first-row diagrams.

We now proceed to compute the one-loop corrections induced by the semi-infinite Wilson lines. The contribution of diagram $(a)$ in Fig.~\ref{fig:Og} takes the form
\begin{eqnarray}
I_a^{(\delta_-)} &=&-g_s^2\frac{C_A}{2}\mu^{2\varepsilon}\frac{\e^{\varepsilon\gamma_E}}{(4\pi)^\varepsilon}\int^0_{-\infty}\df \lambda\,\int\frac{\df^Dp}{(2\pi)^D}\int\frac{\df^Dl}{(2\pi)^D}\e^{-i n_-\cdot[\lambda(l+p)+ tp]-i (\lambda+t)\delta_-}\notag\\
		&&\hspace*{30mm}\times\frac{\bar{q}(p)\slashed{n}_-\slashed{l}}{l^2 (p+l)^2}T^a\frac{\slashed{n}_-\slashed{n}_+}{4}T^bq(sn_+)\notag\\		
		&=& \frac{i\alpha_s(\mu) C_A\e^{\varepsilon\gamma_E}\Gamma(\varepsilon)}{4\pi}\,\int_0^1\df\alpha\,\frac{\alpha^{1-\varepsilon}}{\bar{\alpha}^{\varepsilon}}\int\frac{\df^Dp}{(2\pi)^D}\left(\frac{-p^2}{\mu^2}\right)^{\!-\varepsilon}\,(n_-\cdot p)\notag\\
		&&\hspace*{0mm}\times \!\int_{-\infty}^{0}\!\df \lambda\,\e^{-i(t+\bar\alpha\lambda)n_-\cdot p-i(t+\lambda)\delta_-} \,\bar{q}(p)T^a\frac{\slashed{n}_-\slashed{n}_-}{4}T^bq(sn_+)\, .
\label{eq:OgIcnew}
\end{eqnarray}
One can now keep only the leading $\varepsilon$ term in the integrand as in the photon case to simplify the calculations without changing the UV and rapidity behavior captured by $\varepsilon$ and $\delta_-$, respectively. Thus we obtain
\begin{align}
   \label{eq:OgIa}
   I_a^{(\delta_-)} &= \frac{\alpha_s(\mu)}{4\pi}\frac{C_A}{\varepsilon}\int\frac{\df^Dp}{(2\pi)^D}\e^{-itn_-\cdot p}\left[\ln\frac{\delta_-}{n_-\cdot p}+1+\mathcal{O}(\delta_-)\right]\,\bar{q}(p)T^a\frac{\slashed{n}_-\slashed{n}_+}{4}T^bq(sn_+)\notag\\
		&= \frac{\alpha_s(\mu)}{4\pi}\frac{C_A}{\varepsilon}\left[\ln\frac{-\delta_-}{\mu}-\ln\frac{-i\partial_t}{\mu}+1+\mathcal{O}(\delta_-)\right]O_g(s,t)+\mathcal{O}(\varepsilon^0)\, ,
\end{align}
where $\partial_t=\partial/\partial t$. We will return to this expression later when we discuss how it relates to conformal symmetry. It is important to note that the $\varepsilon,\delta_\pm\to0$ limit must be taken in the order of $\displaystyle\lim_{\varepsilon\to0}\lim_{\delta_\pm\to0}$. The mirror diagram is obtained by changing $-i\partial_t$ to $i\partial_s$ and replacing $-\delta_-$ by $\delta_+$ in the square bracket above. The result in~\eqref{eq:OgIa} remains the same in general $R_\xi$ gauge.

For diagram $(b)$ in Fig.~\ref{fig:Og}, we have 
\begin{align}
	\label{eq:OgIb}
		I_b^{(\delta_-)} &= 2ig_s^2 \mu^{2\varepsilon} \frac{C_A}{2} \,\frac{\e^{\varepsilon\gamma_E}}{(4\pi)^\varepsilon}\int^0_{-\infty}\df \lambda\, \int_0^1\df v\,s \int\frac{\df^Dl}{(2\pi)^D}\frac{\e^{-i l\cdot[(t+\lambda)n_--vs n_+] -i(t+\lambda)\delta_-}}{l^2}\notag\\
		&\hspace*{25mm}\times\bar{q}(tn_-)T^a\frac{\slashed{n}_-\slashed{n}_+}{4}T^bq(sn_+)\notag\\
		&= \frac{\alpha_s(\mu)}{4\pi}C_A\e^{\varepsilon\gamma_E}\Gamma(1-\varepsilon)\,(s\mu^2)^{\varepsilon}\int^0_{-\infty}\df \lambda\int_0^1\df v\,[(\lambda+t)v]^{\varepsilon-1}\,\e^{{-i(\lambda+t)\delta_-}}\notag\\
		&\hspace*{25mm}\times\bar{q}(tn_-)T^a\frac{\slashed{n}_-\slashed{n}_+}{4}T^bq(sn_+)\notag\\
		&= \frac{\alpha_s(\mu)}{4\pi}\frac{C_A}{\varepsilon}\Big[\ln\frac{-\delta_-}{\mu}+\ln\left(-i\mu t \e^{\gamma_E}\right)+\mathcal{O}(\delta_-)\Big]O_g(s,t)+\mathcal{O}(\varepsilon^0)\, .
\end{align}
Its mirror counterpart is obtained similarly to diagram $(a)$ ($\{\delta_-, t \}\mapsto \{-\delta_+, -s\} $). It is worth pointing out that the $\lambda$-integral is well-defined despite the potential complications in the vicinity of $\lambda=-t$ and the branch cut generated by $\lambda+t<0$ thanks to the $+i0^+$ prescription introduced for $t$ necessary to make the Fourier transform of the anomalous dimension invertible (see remarks below~\eqref{eq:softdef}). This prescription moves the variable $(\lambda+t)$ off the real axis into the upper complex plane, thereby circumventing the problematic region of $\lambda\leq -t$. The imaginary unit in $\ln(-i\mu t\e^{\gamma_E})$ of~\eqref{eq:OgIb} is canceled out by the mirror of diagram $(b)$, rendering the total sum proportional to $\ln(\mu^2st)$.\footnote{The phase of $\ln(-i\mu t)$ in the position space anomalous dimension has no physical significance as its momentum-space counterpart remains real-valued. Here, the evolution kernel of the $B$-meson LCDA serves as a primary example. The presence (absence) of the phase in the position (momentum) space anomalous dimension is a consequence of the peculiarity of the Fourier transform in~\eqref{eq:Fourier} with positive momentum support.} Carrying out the calculation in general $R_\xi$ gauge, we find identical results. 

Diagram $(c)$ translates into the loop integral
\begin{align}
    I_c^{(\delta_\mp)} &= -2i g_s^2f^{ace}f^{bde} \mu^{2\varepsilon}\frac{\e^{\varepsilon\gamma_E}}{(4\pi)^\varepsilon} \int^0_{-\infty} \df \lambda_1 \int_{-\infty}^0 \df \lambda_2 \int\frac{\df^Dl}{(2\pi)^D}\frac{\e^{-i l\cdot ((\lambda_1+t)n_- -  (\lambda_2+s) n_+)}}{l^2} \notag\\
    &\hspace*{25mm}\times\e^{-i (\lambda_1+t)\delta_- - i (\lambda_2+s)\delta_+}\bar q(tn_-)T^c\frac{\slashed{n}_-\slashed{n}_+}{4}T^d q(sn_+)\notag\\
   &= g_s^2f^{ace}f^{bde} \mu^{2\varepsilon} \e^{\varepsilon\gamma_E} \int^0_{-\infty} \df \lambda_1 \int_{-\infty}^0 \df \lambda_2\,  \left[-\frac{\Gamma(1-\varepsilon)}{8\pi^2} \big((\lambda_1+t) (\lambda_2+s)\big)^{\varepsilon-1}\right]  \notag \\
   &\hspace*{25mm}\times \e^{-i (\lambda_1+t)\delta_- - i (\lambda_2+s)\delta_+} \bar q(tn_-)T^c\frac{\slashed{n}_-\slashed{n}_+}{4}T^d q(sn_+)\,.
   \label{eq:Icdel}
\end{align}
Technically,  $\lambda_1+t$ ($\lambda_2+s$) is a variable in the upper (lower) half of the complex plane, as explained above, which bypasses the potential singularities at $\lambda_1=-t$ and $\lambda_2=-s$.   
Following the procedure of taking the limit in the order of $\displaystyle\lim_{\varepsilon\to0}\lim_{\delta_\pm\to0}$, we find $I_c$ to be finite. 
From a different perspective, the UV divergences are associated with fluctuations in the small or light-like distances, both of which are absent in diagram $(c)$ as the two vertices connected by the gluon propagator never coincide, nor does their separation become light-like. We therefore immediately conclude that diagram $(c)$ does not contribute to the anomalous dimension. This conclusion remains unchanged when $R_\xi$ gauge is adopted.

The same argument also applies to diagram $(d)$, hence it does not contribute to the RG kernel either. 
To see this, we start from the loop integral
\begin{align}
I_d^{(\delta_-)} 
=\,& -ig_s^2 f^{acd}  \mu^{2\varepsilon}\frac{\e^{\varepsilon\gamma_E}}{(4\pi)^{\varepsilon}}\,\int^0_{-\infty} \!\df \!\lambda\,\e^{-i(\lambda+t) \delta_-} \!\int\frac{\df^Dp}{(2\pi)^D}\int\frac{\df^Dl}{(2\pi)^D}\,\e^{-i[s n_+\cdot l+(\lambda+t)n_-\cdot (p-l)]}
\nonumber\\
&\hspace*{3cm}\times\,\bar{q}(tn_-)T^cT^bT^d\frac{\slashed{n}_-\slashed{n}_+}{4}\frac{\slashed{l}\slashed{n}_-}{l^2 (p-l)^2 }q(p)\notag\\
    =\,& \frac{\alpha_s(\mu)}{\pi}f^{acd} \mu^{2\varepsilon} \e^{\varepsilon\gamma_E} \int\frac{\df^Dp}{(2\pi)^D}\,(n_-\cdot p)\int^0_{-\infty} \df \lambda\int^1_0 \df \alpha\,\alpha\,\e^{-i((\lambda+t)\bar\alpha n_-+\alpha s n_+ )\cdot p}\notag\\
&\times\left(-\frac{(\lambda+t)s}{\alpha\bar\alpha p^2}\right)^{\!\varepsilon/2}\!K_{\varepsilon}\left(2\sqrt{-p^2\alpha\bar\alpha (\lambda+t) s}\right) \bar q(tn_-) T^c T^b \frac{\slashed{n}_-\slashed{n}_+}{4} T^d q(p)\,.
\end{align} 
Since $(\lambda + t)$ is properly defined thanks to the $+i0^+$ prescription for $t$, we can split the integration region as 
\begin{align}
    \int^0_{-\infty} \df \lambda\,\big(\cdots\big) = \int^0_{-t} \df u\,\big(\cdots\big) + \int^{-t}_{-\infty} \df u\,\big(\cdots\big) \equiv {\cal I}_1 + {\cal I}_2
\end{align}
to simplify the calculation for $t>0$.  
We can take $\delta_-=0$ from the start in ${\cal I}_1$ without changing the UV behaviour, which allows us to conclude that ${\cal I}_1$ is UV-finite following~\eqref{eq:BKas}. The contribution of ${\cal I}_2$ can be calculated by a change of variable $\lambda\mapsto -t+\lambda^2/(-4p^2\alpha\bar\alpha s)$ after which the $\lambda$-integral can be performed using the identity
\begin{align}
    \int_0^{\infty} \df z\, z^{1+\varepsilon}\e^{-a z^2}K_\varepsilon(z) = \frac{\e^{\frac{1}{4a}}\Gamma(1+\varepsilon)}{2^{2-\varepsilon}a}\,{\rm E}_{1+\varepsilon}\!\left(\frac{1}{4a}\right)\, ,\hspace*{5mm} a>0,\,\,\varepsilon>-1\, ,
\end{align} 
where ${\rm E}_n(z)$ denotes the exponential-integral function. 
The resulting $\alpha$-integral takes a relatively simple form and an analysis of the integrand, particularly in the endpoint region $\alpha\to 0,1$, finally reveals the UV-finiteness of diagram $(d)$ in the $\displaystyle\lim_{\varepsilon\to0}\lim_{\delta_\pm\to0}$ limit. Furthermore, explicit calculations also show that diagram $(d)$ does not contribute to the renormalization of $O_g$ in general $R_\xi$ gauge.

Summing up all contributions together, we obtain the following one-loop result for the unsubtracted operator $O_g(s,t)$ in general $R_\xi$ gauge:
\begin{align}
\label{eq:Ogpart}
		I_{O_g^{\rm uns}}^{(\delta_\pm,\xi)} =\,&  O^{\rm uns}_g(s,t) +
		\frac{\alpha_s(\mu)}{4\pi}\frac2\varepsilon\left(C_F-\frac{C_A}{2}\right)\int_0^1\df u\left[\frac{u}{\bar u}\right]_+\Big(O_g^{\rm uns}(us, t)+O_g^{\rm uns}(s, ut)\Big)\notag\\
		&+\frac{\alpha_s(\mu)}{4\pi}\left[\frac{C_A}{\varepsilon}\left(2+\ln \left(st\mu^2\e^{2\gamma_E}\right)+2\ln\frac{-\delta_-\delta_+}{\mu^2}-\ln\frac{\partial_s\partial_t}{\mu^2}\right)\right.\notag\\
		&\qquad\left.-2C_F\left(\frac{1}{\varepsilon^2}+\frac{\ln \left(st\mu^2\e^{2\gamma_E}\right)}{\varepsilon} \right)+\frac{C_A}{\varepsilon}(1-\xi)-\frac{C_F}{ \varepsilon}\right]O_g^{\rm uns}(s,t)  \,.
\end{align}
We added the ``tree-level" contribution $O_g(s,t)$ in the above for later convenience. Note that with the inclusion of the quark self-energy, the $\delta_\pm$ and $\xi$ dependence is proportional only to the colour factor $C_A$. The former is a result of interactions involving the adjoint semi-infinite Wilson lines, while the latter originates solely from the third diagram in the first row of Fig.~\ref{fig:Og} contributing with colour factor $C_F-C_A/2$ to the gluon case.

\subsection{Subtracted IR/rapidity-finite \texorpdfstring{$O_g(s,t)$}{}}
\label{sec:subtraction}

The dependence of the UV poles in \eqref{eq:Ogpart} on the IR/rapidity 
regulators $\delta_\pm$ implies that one cannot consistently define an 
IR-finite anomalous dimension for $O_g^{\rm uns}$, unlike for $O_\gamma$. 
The problem arises from the two incoming energetic gluons, which give rise to the semi-infinite adjoint Wilson lines, representing the incoming colour charges 
``from infinity''. The situation is analogous to the factorization of 
QED effects in the rare decay $B_s\to \ell^+\ell^-$~\cite{Beneke:2019slt}, 
where the electric charges moving to infinity cause a similarly ill-defined soft-function anomalous dimension. The remedy is to define a subtracted soft 
function, which rearranges terms between the soft and (anti-) collinear functions of the factorization formula for the process, which cures a similar problem 
in the (anti-) collinear sector. 

For the $gg\to h$ soft function, the subtraction  operator is naturally defined as 
\begin{equation}
	\label{eq:subop}
	\mathbb{S}_g(0) = \big(\mathcal{Y}_{n_-}(0)\big)^{ac} \big(\mathcal{Y}_{n_+}(0)\big)^{cb} =  \big[ -\infty n_-, 0n_-\big]^{ac}\big[0n_+, -\infty n_+\big]^{cb}\,.
\end{equation}
The position $0$ is where the $q\bar{q}h$ interaction happens and the two free colour indices $a$ and $b$ reside at the infinity. The explicit definition of the adjoint Wilson lines in the subtraction operator follows~\eqref{eq:adjW} including the $\delta_\pm$ regulators. We are now ready to define the rapidity-finite soft operator entering the factorization formula for $gg\to h$ process mediated by a light-quark loop as
\begin{align}\label{def:Osb}
    {O}_g(s,t) \equiv \frac{O^{\rm uns}_g(s,t)}{\langle \mathbb{S}_g(0) \rangle }\,,
\end{align}
where $\langle  \mathbb{S}_g(0)\rangle$ denotes the vacuum expectation of $\mathbb{S}_g(0)$ with the trivial colour tensor removed,
\begin{align}
\label{eq:rprm}
    \delta^{ab}\langle  \mathbb{S}_g(0)\rangle \equiv \langle 0 |{\bf T}\,\mathbb{S}_g(0) |0\rangle\rangle \equiv \delta^{ab} R_-^{(\xi)} R_+^{(\xi)} 
\end{align}
with $R_\pm$ defined here for later convenience. The superscript $(\xi)$ emphasizes that $R_\pm^{(\xi)}$ are gauge-\textit{dependent} objects as will be seen below.

\begin{figure}[t]
  \centering
  \includegraphics[width=8cm]{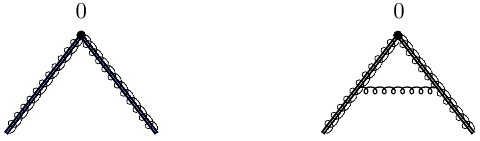}
  \caption{\label{fig:subtraction} Double lines superposed by curly ones represent light-like adjoint Wilson lines. The simple curly line denotes the gluon propagator. The self-energy of the Wilson lines vanishes in $R_\xi$ gauge hence not shown.}
\end{figure}

The tree diagram and the only non-vanishing one-loop diagram are shown in Fig.~\ref{fig:subtraction}. The leading order result is $\langle  \mathbb{S}_g(0) \rangle=1+\mathcal{O}(\alpha_s)$ and the one-loop contribution is quite simple. In $R_\xi$ gauge, we find 
\begin{eqnarray}
		\langle  \mathbb{S}_g(0) \rangle 
		&=& 1 - ig_s^2 C_A 
		\,\frac{\mu^{2\epsilon}\e^{\varepsilon\gamma_E}}{(4\pi)^\varepsilon}\int^0_{-\infty}\df \lambda_1\, 
		\df \lambda_2 \int\frac{\df^Dl}{(2\pi)^D}\left(\frac{2}{l^2}-(1-\xi)\frac{n_-\cdot l\,  n_+\cdot l}{[l^2]^2}\right)\notag\\
		&&\times \,\e^{-i \lambda_1(\delta_--n_-\cdot l) - i \lambda_2(\delta_+-n_+\cdot l)}\notag\\[0.2cm]
		&=& 1 + \frac{\alpha_s(\mu)}{4\pi}C_A\left[-\frac{2}{\varepsilon^2}+\frac{2\ln(-\delta_-\delta_+/\mu^2)+(1-\xi)}{\varepsilon}\right]\, ,
	\label{eq:subtraction}
 \end{eqnarray}
where all divergences in either the $\varepsilon \to 0$ or the $\delta_\pm \to 0$ limit (or both) are kept.\footnote{We point out that the gauge dependence of  $\langle  \mathbb{S}_g(0) \rangle$ using the $\delta$-regulator was already observed in~\cite{Echevarria:2015byo}, stemming from $\delta$ playing the role of an IR regulator  when $\xi \neq 1$ at the one-loop order. 
While, in principle, the $\xi$ dependence can be removed by a manipulation of the integrand~\cite{Echevarria:2015byo}, this would be contrary to the spirit of the subtraction procedure, where the gauge dependence of  $\langle  \mathbb{S}_g(0) \rangle$ is required to cancel the one already present in the unsubtracted soft function. It would be interesting to check whether the complete cancellation of $\delta_\pm$ and $\xi$ dependence between ${\cal I}_{O_g}^{(\xi)}$ and $R_-^{(\xi)}R_+^{(\xi)}$ observed below persists to higher loop orders.
}
Combining with \eqref{eq:Ogpart}, we obtain
\begin{align}
	\label{eq:subtracted}
		{\cal I}_{O_g}^{(\xi)} \equiv \frac{I_{O_g^{\rm uns}}^{(\delta_\pm,\xi)}}{R_-^{(\xi)}R_+^{(\xi)}}=\,& 1+ \frac{\alpha_s(\mu)}{4\pi\varepsilon}\Bigg\{2\left(C_F-\frac{C_A}{2}\right)\int_0^1\df u\left[\frac{u}{\bar u}\right]_+\Big(O_g(us, t)+O_g(s, ut)\Big) \notag\\
		&\hspace*{-1.2cm}+\left[C_A\left(\frac{2}{\varepsilon}+2+\ln \left(st\mu^2\e^{2\gamma_E}\right)\right)-2C_F\left(\frac{1}{\varepsilon}+\ln \left(st\mu^2\e^{2\gamma_E}\right)\right)\right]O_g(s,t) \notag\\
		&\hspace*{-1.2cm}-C_A\ln\frac{\partial_s\partial_t}{\mu^2} O_g(s,t) -C_F O_g(s,t) \Bigg\} +{\cal O}(\alpha_s^2)\, .
	\end{align}
 We see clearly that both the $\xi$ and $\delta_\pm$ dependence are completely removed by the subtraction. Note that after substituting $\alpha_s C_A\to \alpha_{\rm em}Q_\ell^2$, the subtraction factor $R_-^{(\xi)}R_+^{(\xi)}$ takes the same form in Feynman gauge as in~\cite{Beneke:2019slt,Beneke:2020vnb} where the QED counterpart of~\eqref{eq:subop} was constructed. 
  
The one-loop renormalization for the \textit{subtracted} gluon operator $O_g$~\eqref{eq:subop} can now be obtained immediately from~\eqref{eq:subtracted}. In position space, it takes the form
\begin{eqnarray}
 O_g(s, t;\mu) &=& O^{\rm bare}_g(s,t)-\frac{\alpha_s(\mu)}{4\pi}\,\Bigg\{\bigg[\frac{2}{\varepsilon^2}(C_A-C_F) - \frac{2}{\varepsilon}\left(C_F-\frac{C_A}2\right)\ln \left(st\mu^2\e^{2\gamma_E}\right)\nonumber\\
 &&-\,\frac{C_A}{\varepsilon}\ln\left(\frac{\partial_s\partial_t}{\mu^2}\right)-\frac{C_F-2C_A}{\varepsilon}\bigg]{O}^{\rm bare}_g(s,t)
    \label{eq:pos-Zg}\\
 &&   +\,\frac{2}{\varepsilon}\!\left(C_F-\frac{C_A}{2}\right) \int^1_0 \df u \left[\frac{u}{\bar u}\right]_+\!\Big(O^{\rm bare}_g(us, t)+O_g^{\rm bare}(s, ut)\Big)\Bigg\}+{\cal O}(\alpha_s^2)\, .
    \nonumber
\end{eqnarray}
The expression $\ln(\partial_s\partial_t/\mu^2)$ is local in the momentum space, namely $\displaystyle\ln(\ell_-\ell_+/\mu^2)=\ln(w/\mu^2)\equiv L_w$ via~\eqref{eq:softdef}. 
It is then straightforward to find the renormalization kernel in momentum space $Z_g(w,w')$ by applying the Fourier transform identities~\eqref{eq:FT1} and~\eqref{eq:FT2} and following the procedure in appendix~\ref{app:FT}, 
resulting in 
\begin{align}
    \label{eq:Zsg}
	Z_{g}(w,w') =\,& \delta(w-w')+\frac{\alpha_s(\mu)}{4\pi}\Bigg\{\left[2(C_F-C_A)\left(\frac{1}{\varepsilon^2}-\frac{L_w}{\varepsilon}\right)-\frac{3C_F}{\varepsilon}\right]\delta(w-w')\notag\\
	&\hspace*{0mm}-\frac{4}{\varepsilon}\left(C_F-\frac{C_A}{2}\right)w\Gamma(w,w')\Bigg\}+{\cal O}(\alpha_s^2)\, ,
\end{align}
where $\Gamma(w,w')$ is defined in~\eqref{eq:LNK}. 
This expression for the renormalization factor is in agreement with the conjecture in \cite{Liu:2022ajh}.\footnote{Up to a $\beta_0$-term because our operator definition does not include an overall $\alpha_s$ (originating in \cite{Liu:2022ajh} from the collinear gluon emission vertices).} Taking $C_A\to 0$, we recover the renormalization factor~\eqref{eq:Zsp} for the photon case.
Finally, we obtain the anomalous dimensions: In position space, from~\eqref{eq:pos-Zg},
\begin{eqnarray}
    [\gamma_g O_g](s,t) &=& -\frac{\alpha_s(\mu)}{2\pi}\,\bigg\{2\left(C_F-\frac{C_A}{2}\right) \int^1_0\df u\,\left[\frac{u}{\bar u}\right]_+\Big(O_g(us, t)+O_g(s, ut)\Big)\notag\\
&&\hspace*{-2.1cm}-\,\bigg[C_A\ln\left(\frac{\partial_s\partial_t}{\mu^2}\right)+ 2\left(C_F-\frac{C_A}2\right)\ln \left(st\mu^2\e^{2\gamma_E}\right) +(C_F-2C_A)\bigg]{O}_g(s,t)\bigg\}
\,.\qquad\quad
\label{eq:1Lg-pos}
\end{eqnarray}
The momentum-space anomalous dimension follows readily from \eqref{eq:Zsg},
\begin{align}
	\label{eq:AD}
	\gamma_g(w, w') &=  \frac{\alpha_s(\mu)}{\pi}\left[\Big((C_A-C_F)L_w-\frac{3}{2}C_F\Big)\delta(w-w')+(C_A-2C_F) w \Gamma(w,w')\right],\notag\\
	&
\end{align}
We note that the analytical solution to the momentum-space one-loop RGE for $O_g$, governed by $\gamma_g$, has been found in~\cite{Liu:2022ajh}.

\subsection{Factorization formula for \texorpdfstring{$gg\to h$}{}}
\label{sec:fac}

From the perspective of factorization, the need for the additional 
regulators $\delta_\pm$ implies a failure of naive soft-collinear 
factorization. In the presence of the in-coming colour charges 
from $-\infty$, representing the (anti-) collinear initial-state 
gluons, soft emissions require an additional rearrangement 
implemented by dividing the soft operator by the 
vacuum expectation value of the subtraction operator $\mathbb{S}_g(0)$ 
(see~\eqref{def:Osb}). Only with this subtraction does the soft operator 
have a well-defined renormalization-group equation on its own. 
The rearrangement means that the modification of the soft operator 
must be compensated by a corresponding modification of the collinear 
functions in the factorization formula \cite{Beneke:2019slt}.

To implement this rearrangement for the process $gg\to h$, we begin by quoting the factorization formula for the $gg\to h$ form factor with a light quark loop from~\cite{Liu:2022ajh},
\begin{align}
	\label{eq:refact}
		F_{gg}(\mu) &=\overbrace{H_1(\mu)S_1(\mu)}^{T_1(\mu)} +\overbrace{4\int_0^1\frac{\df z}{z}\Big(\bar{H}_2(z,\mu)S_2(z,\mu) -\braces{\bar{H}_2(z,\mu)}\braces{S_2(z,\mu)}\Big)}^{T_2(\mu)}\notag\\
		&+\underbrace{\lim_{\sigma\to -1}H_3(\mu)\int_0^{M_h}\frac{\df\ell_-}{\ell_-}\int_0^{\sigma M_h}\frac{\df\ell_+}{\ell_+}J_g(M_h\ell_-,\mu)J_g(-M_h\ell_+,\mu)S_g(\ell_-\ell_+,\mu)}_{T_3(\mu)} \notag\\
		&\equiv  T_1 + T_2 + H_3 \cdot [J_g(M_h\ell_-)J_g(-M_h\ell_+)] \otimes \widehat{S}_g(\ell_-\ell_+)\, .
\end{align}
The $\otimes$ symbol refers to the convolution in light-cone momentum components $\ell_\pm$. For the following, only the last term, $T_3$, in the formula is relevant. The point is that the soft function $\widehat{S}_g$ is defined here 
as the matrix element of the unsubtracted operator \eqref{eq:Ogdef}, 
which cannot be consistently renormalized on its own. 

The dimensionally regulated, on-shell $gg\to h$ matrix element can be 
written (focusing on the $T_3$ term above) as 
\begin{equation}
\label{eq:gghmatrixelement}
{\cal M}_{gg\to h} \supseteq H_3\cdot \big[J_g(M_h\ell_-)J_g(-M_h\ell_+)\big]\otimes \widehat{S}_g(\ell_-\ell_+) \,
{\mathfrak A}_{g}^{+,(\xi)}(k_1^2)\,{\mathfrak A}_{g}^{-,(\xi)}(k_2^2)\,,
\end{equation}
where the factors ${\mathfrak A}_{g}^{\pm,(\xi)}(k_i^2)$ arise from 
the  (soft-decoupled) collinear matrix element 
of the initial-state gluon(s),
\begin{equation}
\label{eq:OSRHS}
\langle 0| \mathcal{A}^{\mu (0)}_{\perp, n_\pm}(0) |g(k,a)\rangle \equiv g_s T^a \varepsilon^\mu_\perp(k)\,{\mathfrak A}_{g}^{\pm,(\xi)}(k^2) \,.
\end{equation}
For convenience, we strip the matrix element of the factors 
$g_s T^a \varepsilon^\mu_\perp(k)$.  We also understand the factors $H_3$, $\widehat{S}_g$, $J_g$, and ${\mathfrak A}_{g}^{\pm,(\xi)}$ 
as the regulated but unrenormalized hard, soft, and collinear functions. 
With this convention, \eqref{eq:OSRHS} 
coincides with \eqref{eq:refact} for {\em on-shell} gluons, since 
${\mathfrak A}_{g}^{\pm,(\xi)}(k^2)|_{k^2=0}=1$ to all orders, because 
the loop integrals are scaleless. Making the factors ${\mathfrak A}_{g}^{\pm,(\xi)}(k^2)$ explicit is crucial for the discussion. Like $\widehat{S}_g$ 
these factors cannot be consistently UV-renormalized on their own, 
because their anomalous dimension depends on the non-dimensional 
IR regulator employed to extract the UV poles. We further mention 
that the on-shell amplitude ${\cal M}_{gg\to h}$ is UV-finite but IR 
divergent. In dimensional regularization the IR divergence can be removed by multiplying 
the matrix element with $Z_{gg}^{-1}$, 
where $Z_{gg}$ has an alternative interpretation as the UV renormalization 
constant of the SCET gluonic two-jet operator  
\begin{equation}
\label{eq:Ogg}
O_{gg}=\mathcal{A}_{\perp,n_-}^{\mu,a}\mathcal{A}_{\perp\mu,n_+}^{a}(0),
\end{equation}
of the {\em non-decoupled} (anti-) collinear SCET gluon fields 
$\mathcal{A}_{\perp,n_\pm}^\mu=\left[W^\dagger_{n_\pm}iD^\mu_\perp W_{n_\pm}\right]$. The computation of this renormalization factor involves 
collinear, anti-collinear {\em and} soft loops and {\em is} well-defined. In the $\overline{\rm MS}$ scheme \cite{Becher:2009cu}, it is given by\footnote{Due to our definition of $O_{gg}$, which does not include a factor of $1/g_s^2$, there is no $\beta_0/\eps$ in $Z_{gg}$, compared with \cite{Liu:2022ajh}.}
\begin{equation}
    \label{eq:Zgg}
    Z_{g g}=1-\frac{\alpha_s(\mu)}{4 \pi}\left[2 C_A\left(\frac{1}{\varepsilon^2}-\frac{L_h}{\varepsilon}\right)\right]+\mathcal{O}\left(\alpha_s^2\right),\quad\text{with}\quad L_h=\ln\frac{-M_h^2}{\mu^2}\,.
\end{equation}

Before proceeding with the rearrangement, we discuss the (anti-) collinear 
(more precisely (anti-) hardcollinear) functions $J_g(\pm M_h\ell_\mp)$ 
in the factorization formula. Quite generally, the collinear functions are defined as matching coefficients of time-ordered products with the subleading power SCET Lagrangian \cite{Beneke:2019oqx}. For the $T_3$ term, the relevant 
definition of the  (position-space) collinear function reads
\begin{align}
    \label{eq:Jetdef}
     & \,i\int \df^4 z\,\, \textbf{T}\left\{\left[\bar{q}_sY_{n_-}\right](z_-)\left[ \slashed{\mathcal{A}}_{\perp,n_-}^{(0)}W_{n_-}^{(0)\dagger} \xi^{(0)}_{n_-}\right](z), \left[\bar{\xi}^{(0)}_{n_-} W_{n_-}^{(0)}\right](\lambda n_+)\right\} \notag\\
    &= \, 2\pi \int\df s\int \df \left(\frac{n_+\cdot z}{2}\right)\tilde{J}_g\left(\lambda, s, \frac{n_+\cdot z}{2}\right)\left[\bar{q}_sY_{n_-}\right](z_-)\,\slashed{\mathcal{A}}^{(0)}_{\perp,n_-}(sn_+)\frac{\slashed{n}_-}{2},
\end{align}
where $z_-=(n_+\cdot z/2) n_-$, and the superscript ``$(0)$'' denotes the soft-decoupled fields, $q_s$ is the soft quark field which will eventually enter the definition of the soft operator, and $\xi_{n_-}$ stands for the collinear quark field in the $n_-$ direction.\footnote{The above definition coincides with 
variants of the matching equation in \cite{Beneke:2020ibj,Liu:2021mac}.}  
The collinear function can be extracted by sandwiching the collinear part of the above matrix element between a collinear gluon and the vacuum state. Since 
the renormalization of the soft-quark function requires the introduction of a non-dimensional IR regulator $\delta_\pm$, the collinear function must be computed with a consistent regulator as well. 
As pointed out in~\eqref{eq:delta-offshell}, the $\delta_\pm$-regulators are related to the off-shellnesses of the (anti-) collinear gluons through
\begin{equation}
    k_1^2= \delta_+\, n_- \cdot k_1=M_h \delta_+\,,\quad k_2^2= \delta_-\, n_+\cdot k_2=M_h \delta_-\,.
\end{equation}
The technical details of this computation in general $R_\xi$ gauge can be found in the App.~\ref{app:jet}. The main 
result is that the collinear functions $J_g(\pm M_h\ell_\mp)$, defined as above as the matching coefficient from $\text{SCET}_{\rm I}$ onto $\text{SCET}_{\rm II}$, remain the same as the on-shell ones and are, in fact, gauge-invariant even with an external off-shell gluon. To arrive at this result, it is important 
that the matrix element of $\slashed{\mathcal{A}}^{(0)}_{\perp,n_-}(sn_+)$, 
which coincides with \eqref{eq:OSRHS} up to a phase factor, and the Dirac matrix, is not unity 
for an off-shell gluon, but involves 
(see~\eqref{eq:J-RHS} for the $\varepsilon$-exact expression)
\begin{equation}
\label{eq:OfSRHS}
{\mathfrak A}_g^{\pm, (\xi)}(k^2) = 1+\frac{\alpha_s(\mu) C_A}{4\pi}\left(\frac{2}{\eps^2}-\frac{2}{\eps}\ln\frac{-k^2}{\mu^2}-\frac{1-\xi}{2\eps}\right)
\end{equation}
in background field $R_\xi$ gauge. The dependence of this expression on the 
off-shellness $k^2$ and gauge-fixing parameter $\xi$ is exactly what is 
required to make the collinear functions gauge-independent and finite 
as $k^2\to 0$.

We are now in the position to return to \eqref{eq:gghmatrixelement} when 
the matrix element is computed with a small off-shellness $k_i^2\ll M_h^2$ of the external gluons. As said, $J_g(\pm M_h\ell_\mp)$ remain unaltered, but 
$\widehat{S}_g$ turns into the unsubtracted soft function $S_g^{\rm uns,(\xi)}$, 
which depends on the $\delta_\pm$-regulators as well as on $\xi$.  
On the other hand, from \eqref{def:Osb}, \eqref{eq:rprm}, we have
\begin{equation}
   \label{eq:Sg}
   S_g(\ell_+\ell_-) = \frac{S_g^{\rm uns, (\xi)}(\ell_-\ell_+)}{R_+^{(\xi)} R_-^{(\xi)}}\, ,
\end{equation}
and from \eqref{eq:subtraction} we can separate $\langle  \mathbb{S}_g(0)\rangle$ into $R_+^{(\xi)}$ and $R_-^{(\xi)}$ in a symmetric way as\footnote{When axial gauges such as the light-cone gauge are used, $R_+$ and $R_-$ have to be \textit{defined} in an asymmetric way such that the collinear and anti-collinear sectors are separately $\xi$ and $\delta$ independent. This is natural as light-cone gauge itself breaks the $+\leftrightarrow -$ symmetry.}
\begin{equation}
   \label{eq:Rpm}
   \begin{aligned}
       R_\pm^{(\xi)} &= 1+\frac{\alpha_s(\mu)}{4\pi}C_A\left[-\frac{1}{\varepsilon^2}+\frac{2}{\varepsilon}\ln\frac{\pm\delta_\pm}{\mu}+\frac{1-\xi}{2\eps}\right]\,.
   \end{aligned}
\end{equation}
This allows us to rewrite \eqref{eq:gghmatrixelement} in the form 
\begin{eqnarray}
{\cal M}_{gg\to h} &\supseteq& H_3\cdot \big[J_g(M_h\ell_-)J_g(-M_h\ell_+)\big]\otimes S_g(\ell_-\ell_+) \nonumber\\[0.1cm]
&&\times (R_+^{(\xi)}\,{\mathfrak A}_{g}^{+,(\xi)}(k_1^2))\,
(R_-^{(\xi)}\,{\mathfrak A}_{g}^{-,(\xi)}(k_2^2))\,,
\label{eq:gghmatrixelementoffshell}
\end{eqnarray}
where now the soft function is independent of the $\delta$-regulator and 
$\xi$, and {\em can} be renormalized on its own. The subtraction factors 
are multiplied back to the external-collinear gluon factors. Combining 
\eqref{eq:Rpm} with \eqref{eq:OfSRHS} and making use of 
$k_1^2=M_h \delta_-$, $k^2_2 = M_h \delta_-$, 
we find that
\begin{equation}
\label{eq:consistency}
R_\pm^{(\xi)}\,{\mathfrak A}_{g}^{\pm, (\xi)}(k_i^2) 
= 1+\frac{\alpha_s(\mu)}{4\pi} C_A \left[\frac{1}{\eps^2}-\frac{2}{\eps}\ln\frac{\mp M_h}{\mu}\right],
\end{equation}
where the gauge and off-shell regulator dependence also cancels exactly. 
Thus the rearrangement not only allows to UV-renormalize the soft function 
on its own, but also the external collinear factors, implementing a restoration of soft-collinear factorization in very much the same way as in 
\cite{Beneke:2019slt,Beneke:2020vnb} for external fermions.

Incidentally, the second line of \eqref{eq:gghmatrixelementoffshell} 
equals 
\begin{equation}
(R_+^{(\xi)}\,{\mathfrak A}_{g}^{+,(\xi)}(k_1^2))\,
(R_-^{(\xi)}\,{\mathfrak A}_{g}^{-,(\xi)}(k_2^2)) = 
 1+\frac{\alpha_s(\mu)}{4\pi} C_A \left[\frac{2}{\eps^2}-\frac{2L_h}{\eps}\right]=Z_{gg}^{-1}\,.
\end{equation}
These poles are removed by the UV renormalization of the external-collinear 
factors.

\section{Conformal symmetry and higher-order results}
\label{sec:CS}

In this section, we demonstrate the power of conformal symmetry in promoting the anomalous dimensions of the soft operators $O_\gamma$ and $O_g$ to higher orders. In particular, we establish an all-order relation between $\gamma_{\gamma}$, the anomalous dimension of $O_\gamma$, and the evolution kernel of the leading-twist non-local heavy-light quark operator. In addition, we construct a general ansatz for the anomalous dimension of $O_g$ at higher orders in $\alpha_s$, test it against the RG consistency using known two-loop hard and gluon jet 
functions.  
Another major advantage of the conformal symmetry technique is that it offers a systematic approach to solving the RGEs and therefrom reveals the asymptotic behaviours of the soft functions, which we discuss in App.~\ref{app:conformal}. 

We start by recalling that the conformal transformations in the light-like directions form an $SL(2,\mathbb{R})$ group generated by the so-called collinear conformal generators. The $n_+$-collinear conformal generators can be written explicitly as~\cite{Braun:2003rp},
\begin{equation}
	\label{eq:generator}
	\widehat{S}_{+}=s^{2} \partial_{s}+2j s
	=s\theta_s+2j s, 
	\quad \widehat{S}_{0}= %
	s\partial_{s}+j
	=\theta_s+j,
	\quad \widehat{S}_{-}=-\partial_{s}\, ,
\end{equation}
where $j=1$ is the conformal spin of the soft quark and 
$\theta_s=s\partial_s$ the Euler operator. The generators of $n_-$-collinear conformal transformations are written analogously as
\begin{equation}
	\label{eq:generatorT}
	\widehat{T}_{+}=t^{2} \partial_{t}+2j t=t\theta_t+2jt, \quad \widehat{T}_{0} = t \partial_{t}+j=\theta_t+j, \quad \widehat{T}_{-}=-\partial_{t}\, .
\end{equation}
These operators are induced by projecting the $D$-dimensional conformal group generators onto the $n_\pm$ light-like directions. The commutation relations read,
\begin{equation}
	\label{eq:commutations}
	\left[\widehat{S}_{+}, \widehat{S}_{-}\right]=2 \widehat{S}_{0}, \quad\left[\widehat{S}_{0}, \widehat{S}_{ \pm}\right]= \pm \widehat{S}_{ \pm}\,;\quad \left[\widehat{T}_{+}, \widehat{T}_{-}\right]=2 \widehat{T}_{0}, \quad\left[\widehat{T}_{0}, \widehat{T}_{ \pm}\right]= \pm \widehat{T}_{ \pm}\,,
\end{equation}
while $\left[\widehat{S}_i, \widehat{T}_j\right] = 0$ for $i, j\in\{0, \pm\}$. Clearly, these operators form a direct product of two $SL(2,{\mathbb R})$ algebras.

At the one-loop order, the anomalous dimension of the soft operators $O_\gamma$ and $O_g$ in~\eqref{eq:1Lgamma-pos} and~\eqref{eq:1Lg-pos} can be expressed equivalently in terms of these conformal operators in the compact form~\cite{Belitsky:2014rba,Braun:2014owa}, 
\begin{align}
    \gamma_\gamma &=  \frac{\alpha_s(\mu)C_F}{\pi}\,\bigg[\ln\left(\mu^2 \widehat{S}_+ \widehat{T}_+\right)+4\gamma_E -\frac32\bigg]\, ,\label{eq:1LOgamma-CR}\\[0.2cm] 
	\gamma_g &=  \frac{\alpha_s(\mu)}{4\pi}\,\bigg[4\left(C_F-\frac{C_A}{2}\right)\Big(\ln\left(\mu^2 \widehat{S}_+ \widehat{T}_+\right)+4\gamma_E-2\Big)+2C_A\bigg(\ln\frac{\widehat{S}_-\widehat{T}_-}{\mu^2}-2\bigg)\notag\\
	&\hspace*{22mm}+2C_F\bigg]\, .\label{eq:1LOg-CR}
\end{align}
The simplicity of the anomalous dimensions in the conformal representation shown above motivates us to explore the application of conformal symmetry techniques beyond the leading order in the subsequent sections.

\subsection{Relating evolution kernels of soft operators and the \texorpdfstring{$B$}--LCDA}\label{sec:GP}

Extending the one-loop results \eqref{eq:ADp}, \eqref{eq:AD}, the position-space anomalous dimensions for ${O}_\gamma(s,t)$ and ${ O}_g(s,t)$, which we denote collectively as $\gamma_S$, can be tentatively written in the position-space representation to all orders in perturbation theory as
\begin{eqnarray}
    [\gamma_S O_S](s,t) &=& \Gamma_{\rm cusp}^{\rm S}(\alpha_s)\int^1_0\df\alpha\,\frac{\omega_S(\alpha_s,\alpha)}{\bar\alpha} \,\Big(2  O_S(s,t)-O_S(\alpha s,t)-O_S(s,\alpha t) \Big) \nonumber\\
    &&\hspace*{-2cm}+\left[\Gamma^{\rm S}_{\rm cusp}(\alpha_s)\ln(s t \mu^2 \e^{2\gamma_E})+\frac12\Gamma_{\rm cusp}^{\rm A}(\alpha_s)\ln\left(\frac{\partial_t\partial_s}{\mu^2}\right)+ 2\widetilde\Gamma_S(\alpha_s) \right]O_S(s, t)\, ,
    \label{eq:gintK}
\end{eqnarray}
where the $s\leftrightarrow t$ symmetry has been implemented. The function $\omega_S(\alpha_s,\alpha)$ ($S=\gamma,g$) has no power-like divergences at the end-points, and $\widetilde\Gamma_S(\alpha_s)$ is a constant.  $\Gamma^{g}_{\rm cusp}\equiv \Gamma_{\rm cusp}^{\rm F} -\Gamma_{\rm cusp}^{\rm A}/2$ and  $\Gamma^{\gamma}_{\rm cusp}\equiv \Gamma_{\rm cusp}^{\rm F}$ where ``F'' (``A") stands for the fundamental (adjoint) representation.\footnote{Note that $\Gamma_{\rm cusp}^{\rm A}$ term in \eqref{eq:gintK} only appears for $S=g$.} The appearance of the cusp anomalous dimension is a consequence of the universality of IR divergences in QCD. 
By analysis of the Feynman diagrams and requiring that $\gamma_S$ is compatible with total space-time translations, the anomalous dimension can in principle have an additional contribution of the form,
\begin{align}
    [\widetilde\gamma_S O_S](s,t) =\int^1_0\df\alpha\int^{\bar\alpha}_0\df\beta\,\chi(\alpha_s;\alpha,\beta)\, O_S(\bar\alpha sn_++\alpha t n_-, \bar\beta t n_-+\beta sn_+)\, ,
\end{align}
which, at the one-loop order, could come from diagram $(d)$ in Fig.~\ref{fig:Ogamma}.
Such a contribution, however, necessarily moves the two soft quarks off the light cone, contradicting the fact that soft interactions do not alter the (anti-) collinear directions of the underlying fields.
From another perspective, such contributions will lead to momentum-space anomalous dimensions depending on $\ell_+$ and $\ell_-$ separately instead of only their product $w=\ell_+\ell_-$, as can be clearly seen from the Fourier transform detailed in App.~\ref{app:FT}, violating boost invariance of the $gg\to h$ decay process. 
Hence, we conclude that the anomalous dimension for both $O_\gamma$ and $O_g$ take the form~\eqref{eq:gintK} to all orders. Namely, we can write,
\begin{align}\label{eq:AM-s;t}
   \gamma_{\gamma} (s,t) =  \gamma_{\gamma}(s,\mu) + \gamma_{\gamma} (t,\mu)\, ,\qquad \qquad
   \gamma_{g} (s,t) =  \gamma_{g}(s,\mu) + \gamma_{g} (t,\mu)\, ,
\end{align}
which allows us to apply the collinear conformal symmetry technique to the collinear and anti-collinear sectors independently. We note that the form of the all-order anomalous dimension for $O_\gamma$ and $O_g$ in~\eqref{eq:gintK} preserves the location of the Higgs production vertex $h\bar qq$ conveniently set at 0 where the two light-like Wilson lines connect. This is similar to the leading-twist $B$-meson LCDA evolution, under which the location of the heavy quark is also fixed at the origin.

Consider the anomalous dimension of $O_\gamma$ at the critical point of $D$-dimensional QCD with the critical coupling $\alpha_s^*$ such that $\beta(\alpha_s^*)=-2\alpha_s^*(\varepsilon+\widetilde\beta(\alpha_s^*))=0$~\cite{Braun:2013tva} where $\widetilde\beta(\alpha_s)$ is the QCD $\beta$-function in four dimensions. First, we note that the full special conformal generator $K^\mu$ acting on $O_\gamma$ receives quantum corrections,
\begin{align}\label{eq:K-split}
    K^\mu(\alpha_s^*; s,t) &= \frac{n_+^\mu}{2} {\cal K}(\alpha_s^*; s)+\frac{n_-^\mu}{2} {\cal K}(\alpha_s^*; t) \notag\\
    &= \frac{n_+^\mu}{2} \left(\widehat S_+ + \widehat{\Delta K}(\alpha_s^*;s)\right) + \frac{n_-^\mu}{2} \left(\widehat T_+ + \widehat{\Delta K}(\alpha_s^*;t) \right)\, ,
\end{align}
where we have applied the $s n_-\leftrightarrow t n_+$ symmetry. Clearly, the tree-level special collinear conformal generators $\widehat S_+$ and $\widehat T_+$ are induced by $n_\mp\cdot K$, respectively.
Similarly, the dilatation operator of the collinear conformal group acting on $O_\gamma$ takes the form
\begin{align}\label{eq:D-split}
    {\cal D}(\alpha_s^*;s, t) = \left[\widehat S_0 +\widehat{\Delta D}(\alpha_s^*; s)\right] +  \left[\widehat T_0 +\widehat{\Delta D}(\alpha_s^*; t)\right] \, ,
\end{align}
with the $\alpha_s^*$ dependence of $\widehat{\Delta D}(\alpha_s^*,s)$ fully determined by the anomalous dimension $\gamma_{\gamma}(s,\mu)$ and the QCD $\beta$-function. This is in contrast to $\widehat{\Delta K}(\alpha_s^*; s)$, which receives additional contributions from the so-called ``conformal anomaly" that can be calculated systematically through a diagrammatic approach~\cite{Braun:2016qlg}.
Since the conformal symmetry at the critical point is \textit{exact}, we can consequently establish the following exact commutation relations~\cite{Korchemsky:1987wg,Braun:2019wyx} using~\eqref{eq:AM-s;t}-\eqref{eq:D-split},
\begin{align}\label{eq:comm-ph}
    [{\cal K}(\alpha_s^*;s),\gamma_\gamma(\alpha_s^*;s,\mu)] = 0\, ,\qquad [{\cal D}(\alpha_s^*;s),\gamma_\gamma(\alpha_s^*;s,\mu)] = \Gamma_{\rm cusp}^{\rm F}(\alpha_s^*)\, .
\end{align}
The commutation relations~\eqref{eq:comm-ph} are readily solved~\cite{Braun:2019wyx} yielding the solution, 
\begin{align}
    \gamma_{\gamma} (s,\mu) = \Gamma_{\rm cusp}^{\rm F}(\alpha_s) \ln \left( \mathcal{K}(\alpha_s;s) \mu \e^{2\gamma_{E}}\right)+ \Gamma_{\gamma}(\alpha_s),\label{eq:Skernel-gamma}
\end{align}
where we replaced $\alpha_s^*$ by an arbitrary $\alpha_s$ and traded the $\varepsilon$-dependence with the ordinary QCD $\beta$-function to obtain $\gamma_\gamma(s,\mu)$ in four-dimensions thanks to the properties of MS-like schemes~\cite{Braun:2013tva}. 

We can further identify ${\cal K}(\alpha_s; z)$ and ${\cal D}(\alpha_s; z)$ as the special conformal and dilatation generator acting on the leading-twist heavy-light operator~\cite{Braun:2019wyx} whose matrix element defines the leading-twist $B$-meson LCDA. The reason 
is that ${\cal K}(\alpha_s;s)$ and ${\cal D}(\alpha_s;s)$ can be systematically computed by considering their UV divergences when acting on the gauge-invariant operator $\bar X(0)[0,sn_+]q(sn_+)$ 
from which the commutation relations~\eqref{eq:comm-ph} follow. 
Here $\bar X(0)$ must be built of fundamental QCD objects such as the quark field, gluon field strength tensor, or a Wilson line so that definite conformal properties apply. Since each local light field ($q, G_{\mu\nu}$) is associated with a conformal generator~\cite{Braun:2003rp} which depends on the location of the light field,  
$\bar X(0)$ can only be a non-local object, i.e., a Wilson line. On the one hand, \eqref{eq:comm-ph} implies that the special conformal symmetry is preserved only in one particular direction $n_+$ or $n_-$, on the other hand, 
the dilatation symmetry is broken by the cusp term $\Gamma_{\rm cusp}$. This implies 
that the Wilson line, which we still denote as $\bar X(0)$, must lie in a direction $v^\mu$ different from $n_+^\mu$, hence forming a cusp with the Wilson line $[0,sn_+]$, and extend to infinity to ensure gauge invariance. In principle, no further constraints can be imposed on $v^\mu$ from the perspective of conformal symmetry. But as we have explicitly shown in Sec.~\ref{sec:Og}, light-like Wilson lines induce unnecessary complications in the form of rapidity divergences. Furthermore, space- and time-like Wilson lines are equivalent for our consideration. Therefore, we are free to choose $v^\mu$ to be a time-like vector fixing $\bar X(0) = [-\infty v,0]$ with $v^2=1$. We can further identify $\bar X(0)$ 
as the heavy quark field $\bar h_v(0)$ in the heavy-quark effective theory (HQET), i.e.,  $\bar X(0) = \bar h_v(0)$, because $\bar h_v(0)=\bar Q(-\infty v) [-\infty v,0]$ with $\bar Q$ being a sterile scalar field at infinity decoupled from the QCD interactions~\cite{Eichten:1980mw}. It is now evident that the collinear conformal generators (anomalous dimension)  associated with $O_B=\bar h_v(0)[0,sn_+]q(sn_+)$ coincide  with~\eqref{eq:generator} (\eqref{eq:1LOgamma-CR})  at the one-loop order up to a constant that cannot be determined from symmetry considerations. This serves as a consistency check. We can thus make full use of the known results for the anomalous dimension and conformal generators of $O_B$~\cite{Braun:2014owa,Braun:2019wyx}.

Comparing to the evolution kernel of the operator $\bar h_v(0)[0,sn_+]q(sn_+)$~\cite{Braun:2019wyx},
\begin{align}\label{eq:B-kernel}
    {\cal H}_B(z,\mu) = \Gamma_{\rm cusp}^{\rm F}(\alpha_s) \ln \left(i \mathcal{K}(\alpha_s; z) \mu \e^{2\gamma_{E}}\right) + \Gamma_+(\alpha_s),
\end{align}
we see that $\gamma_{\gamma} (s,t,\mu)$ and ${\cal H}_B(s,\mu) + {\cal H}_B^*(t,\mu)$, with ``${}^*$" denoting the complex conjugate, may differ by the constant $2(\Gamma_\gamma(\alpha_s)-\Gamma_+(\alpha_s))$. It can be fixed by considering the local limit $s,t\to 0$ of the identity 
\begin{align}
2\Gamma_\gamma(\alpha_s) - 2\Gamma_+(\alpha_s)  = \gamma_{\gamma} (s,t,\mu)-({\cal H}_B(s,\mu) + {\cal H}^*_B(t,\mu))\, .\label{eq:KS-KB}
\end{align}
This difference removes the locally divergent $\Gamma_{\rm cusp}\ln(\tilde\mu^2 st)$ term in the position-space  representation, making the limit well-defined.  While the remaining terms in $\gamma_{\gamma} (s,t,\mu)$ approach $2\gamma_q$, the non-cusp part of the quark collinear  anomalous dimension~\cite{Matsuura:1988sm},  
${\cal H}_B(s,\mu) + {\cal H}_B^*(t,\mu)$ becomes a double-copy of the anomalous dimension $2(\gamma_q+\gamma_Q)$ that appears for the heavy-to-light operator. 
We therefore conclude that\footnote{While it is tempting to arrive at a stronger conclusion $\Gamma_+ = \gamma_q+\gamma_Q$ and $\Gamma_\gamma = \gamma_q$, this is  incorrect. The reason is that the non-constant part of the anomalous dimension $\ln \left(i \mathcal{K}(\alpha_s) \mu \e^{2\gamma_{E}}\right)$ also generates a non-trivial constant in the local limit even after the singular term $\Gamma_{\rm cusp}\ln(i\mu \e^{\gamma_E} s)$ has been removed. Such contributions are, however, canceled out in~\eqref{eq:KS-KB}. } $2\Gamma_\gamma(\alpha_s) - 2\Gamma_+(\alpha_s) = 2\gamma_q-2(\gamma_q+\gamma_Q) = -2\gamma_Q$, hence 
\begin{align}\label{eq:O_ga-const}
\Gamma_\gamma(\alpha_s) = \Gamma_+ -\gamma_Q\, .
\end{align}
Putting everything together, we find,
\begin{align}\label{eq:gamma_full}
    \gamma_{\gamma} (s, t) &= \Gamma_{\rm cusp}^{\rm F}(\alpha_s) \left[\ln \left( \mathcal{K}(\alpha_s;s) \mu \e^{2\gamma_{E}}\right) + \ln \left( \mathcal{K}(\alpha_s;t) \mu \e^{2\gamma_{E}}\right)\right]+ 2\Gamma_+ -2\gamma_Q  \notag\\[0.2cm]
    &={\cal H}_B(s,\mu) + {\cal H}_B(t, \mu) - 2\gamma_Q\, ,
\end{align}
which is valid to all orders in perturbation theory. 

As an application of~\eqref{eq:gamma_full}, we obtain the two-loop $\mathcal{O}(\alpha_s^2)$ correction to the anomalous dimension $\gamma_\gamma(s,t)$ by first fixing the constant part. The explicit expression for $\Gamma_+$ in~\eqref{eq:B-kernel} is known to the two-loop order and reads~\cite{Braun:2019wyx}
\begin{eqnarray}
    \Gamma_+ &=& \frac{\alpha_s}{4\pi}(-5C_F) + \left(\frac{\alpha_s}{4\pi}\right)^2C_F\,\bigg\{C_F\left[-\frac32+2\pi^2-24\zeta_3\right]+C_A\left[\frac{67}{54}-\frac{43}{18}\pi^2-6\zeta_3\right]\notag\\
    &&\hspace*{0mm}+\,2n_f T_F\left[\frac{13}{27}+\frac{5}{9}\pi^2\right]\bigg\} +\mathcal{O}(\alpha_s^3)\, ,
    \label{eq:GammaPlus}
\end{eqnarray}
where $T_F=1/2$ is the trace of the $SU(N_c)$ generators in the fundamental representation and $n_f$ is the number of active flavors. The heavy-quark anomalous dimension $\gamma_Q$ at two loops reads~\cite{Becher:2009kw}
\begin{align}
    \gamma_Q =\frac{\alpha_s}{4\pi}\left(-2C_F\right) + \left(\frac{\alpha_s}{4\pi}\right)^2C_F\left[C_A\left(\frac{2\pi^2}{3}-\frac{98}9-4\zeta_3\right)+n_f T_F\frac{40}{9}\right] + \mathcal{O}(\alpha_s^3) \, .
\end{align}
Then from~\eqref{eq:O_ga-const}, the constant $2\Gamma_\gamma$ entering the conformal representation of the anomalous dimension $\gamma_\gamma(s,t)$ is 
\begin{eqnarray}
    2\Gamma_\gamma &=&\frac{\alpha_s}{4\pi}(-6C_F) + \left(\frac{\alpha_s}{4\pi}\right)^2C_F\bigg\{C_F\left[-3+4\pi^2-48\zeta_3\right]+C_A\left[\frac{655}{27}-\frac{55\pi^2}{9}-4\zeta_3\right]\notag\\[0.1cm]
    &&\hspace*{00mm} +\, 4n_fT_F\left[-\frac{47}{27}+\frac{5\pi^2}{9}\right]\bigg\} +{\cal O}(\alpha_s^3)\, .
\label{eq:2G_g}\end{eqnarray}
The full expression for the anomalous dimension in the conformal representation then follows from~\eqref{eq:Skernel-gamma},
\begin{align}
    \gamma_\gamma(s,t) &= \Gamma_{\rm cusp}^{\rm F}(\alpha_s) \left[\ln \left( \mathcal{K}(\alpha_s;s) \mu \e^{2\gamma_{E}}\right) + \ln \left( \mathcal{K}(\alpha_s;t) \mu \e^{2\gamma_{E}}\right)\right]+ 2\Gamma_{\gamma}(\alpha_s)\, .
\label{eq:fullgammagam}
\end{align}
The cusp anomalous dimension to the two-loop order reads~\cite{Korchemsky:1987wg},
\begin{align}
    \Gamma_{\rm cusp}^{\rm F}&=\frac{\alpha_s}{4\pi}\left(4C_F\right)+\left(\frac{\alpha_s}{4\pi}\right)^2C_F\left\{C_A\left(\frac{268}{9}-\frac{4\pi^2}{3}\right)-\frac{80}{9}n_fT_F\right\}+{\cal O}(\alpha_s^3)\,, 
\end{align}
and the $\mathcal{O}(\alpha_s)$ correction to the special conformal generator ${\cal K}$ is available from~\cite{Braun:2019wyx},
\begin{align}
    {\cal K}(\alpha_s;s) O(s) &= \widehat S_+ O(s)   + \frac{\alpha_s}{4\pi}s\,\Bigg\{\left(\frac{11}{3}C_A-\frac{4}{3}n_f T_F\right)  O(s) + C_F\bigg[3O(s) \notag\\   
    &\hspace*{0mm}+2\int^1_0\df u\,\left(2\frac{u}{\bar u}+\ln \bar u\right)[O(s) -O(u s) ]\bigg] \Bigg\} + {\cal O}(\alpha_s^2)\,,
\end{align}
with  $O(s)$ being a test function. 

Similarly, the constant term $\widetilde\Gamma_\gamma$ in the position-space representation~\eqref{eq:gintK} 
(with $S=\gamma$) of $\gamma_\gamma(s,t)$ can be computed using 
$\widetilde\Gamma_\gamma  = \gamma_+ - \gamma_Q$
with $\gamma_+$ given explicitly in Eq.~(27) of~\cite{Braun:2019wyx} yielding,
\begin{eqnarray}
        2\widetilde\Gamma_\gamma &=& \frac{\alpha_s}{4\pi}(2C_F) 
    + \left(\frac{\alpha_s}{4\pi}\right)^2C_F\,\bigg\{4C_F\left[\frac{21}{4}+\frac{2\pi^2}{3}-12\zeta_3\right]+C_A\left[\frac{1471}{27}-\frac{35\pi^2}{9}-4\zeta_3\right]\notag\\
    &&\hspace*{00mm}+\,4n_fT_F\left[-\frac{95}{27}+\frac{\pi^2}9\right]\bigg\}\, .\label{eq:2tGg}
\end{eqnarray}
The anomalous dimension in position-space then takes the form
\begin{align}\label{eq:g_g2L}
    [\gamma_\gamma O_\gamma](s,t) &= -\Gamma_{\rm cusp}^{\rm F}(\alpha_s)\bigg\{\int^1_0\df u\,\left[\frac{u}{\bar u}\left(1+\frac{\alpha_s}{4\pi}h(u)\right)\right]_+(O_\gamma(us,t)+O_\gamma(s,u t))\notag\\[0.2cm]
&\hspace*{00mm}-\ln( s t\tilde\mu^2)O_\gamma(s,t)\bigg\} + 2\widetilde\Gamma_\gamma
\end{align}
 at two loops, where 
\begin{align}
        h(u) &=\ln u\left[\left(\frac{11}{3}C_A-\frac{4}{3}n_f T_F\right)+2C_F\left(\ln u-\frac{1+u}{u}\ln\bar u-\frac32\right)\right].
\end{align}

We confirmed that the Fourier transform of~\eqref{eq:g_g2L} reproduces the two-loop momentum-space anomalous dimension $\gamma_{\gamma}$ reported in~\cite{Liu:2020eqe}, where it was extracted from RG consistency of the $\gamma\gamma\to h$ factorization formula. We note that the two-loop expressions of $\gamma_{\gamma}$, which participates in the light-quark induced $\gamma\gamma\to h$ process at the three-loop order (i.e., to ${\cal O}(\alpha_{\rm em}\alpha_s^2)$), has also been verified by comparing the predicted large logarithms with numerical calculations~\cite{Liu:2020wbn}. 

It is also interesting to point out that we expect the Casimir scaling for $O_\gamma$ to hold 
 at least to the two-loop order. Namely, by having the soft fields and the two finite-distance Wilson lines 
in \eqref{def:Oga} in an  
arbitrary representation R of $SU(N_c)$,\footnote{It is understood that the two soft quarks fields $\bar q_s(tn_-)$ and $q_s(sn_+)$ in $O_\gamma^{\rm R}$ are decoupled from the QCD Lagrangian and the active quarks participating in the gluon field renormalization are still in the fundamental representation.} the anomalous dimension $\gamma_\gamma^{\rm R}$ of the resulting operator $O_\gamma^{\rm R}$ can be obtained by simply replacing $C_F \mapsto C_R$ in~\eqref{eq:fullgammagam} and~\eqref{eq:g_g2L}, where $C_R$ denotes the Casimir value of representation R. 
This follows from the well-known Casimir scaling of the cusp anomalous dimension,\footnote{Casimir scaling is broken for the cusp anomalous dimension from the four-loop order~\cite{Moch:2004pa,Henn:2019swt,vonManteuffel:2020vjv}.} 
the Casimir scaling of the heavy-quark anomalous dimension  at this order \cite{Beneke:2009rj}, the trivial change of colour factors in the one-loop correction to the conformal operator ${\cal K}$ in~\eqref{eq:gamma_full}, and the Casimir scaling for $\Gamma_+$ corresponding to $C_F\mapsto C_R$ in~\eqref{eq:GammaPlus}.

\subsection{Bootstrap the two-loop anomalous dimensions of \texorpdfstring{$O_g$}{Lg}}
\label{sect:bootstrap}

In this subsection, we investigate the possibility of bootstrapping the anomalous dimension of $O_g$ to higher orders without explicit calculations. 
It turns out that, despite the additional $\widehat{S}_-$ and $\widehat{T}_-$ contributions in comparison with the case of $O_\gamma$ (see~\eqref{eq:1LOgamma-CR} and \eqref{eq:1LOg-CR}), it is still feasible to some extent.  
First we note that $\widehat{S}_-$ and $\widehat{T}_-$ are induced by the adjoint Wilson line extending to infinity and their appearance is a direct consequence of the rapidity divergence in the original $O_g^{\rm uns}$. The subsequent introduction of the subtraction operator~\eqref{eq:subop} brings extra constants to $\gamma_g$ as well. To be more precise, the first commutation relation in~\eqref{eq:comm-ph} becomes invalid, and we can only expect a single commutator identity, which we conjecture to hold at least to two loops with the conformal operators unchanged from the $O_\gamma$ case:
\begin{align}\label{eq:Kg-comm}
       [{\cal D}(\alpha_s^*),\gamma_g(\alpha_s^*;s,\mu)] = \Gamma_{\rm cusp}^{\rm F}(\alpha_s^*) -  \Gamma_{\rm cusp}^{\rm A}(\alpha_s^*) \, .
\end{align}
Here, $\Gamma^{\rm A}_{\rm cusp}$ reflects the presence of adjoint Wilson lines. In this case, the anomalous dimension $\gamma_g(\alpha_s^*;s,\mu)$ can no longer be solved for unambiguously, and we resort to the general ansatz 
\begin{align}
    \gamma_g (s,\mu) &= \left(\Gamma_{\rm cusp}^{\rm F}(\alpha_s)-\frac12\Gamma_{\rm cusp}^{\rm A}(\alpha_s)\right) \ln \left(i \mathcal{K}(\alpha_s;s) \mu \e^{2\gamma_{E}}\right)\notag\\ 
    &+ \frac12\Gamma_{\rm cusp}^{\rm A}(\alpha_s)\ln\left(\frac{\widehat S_-}{\mu}\right)+ \Gamma_{g}(\alpha_s)\,,
    \label{eq:gKernel-ansatz}
\end{align}
which is motivated by the one-loop result, satisfies the commutation relation~\eqref{eq:Kg-comm}, and takes the form of~\eqref{eq:gintK} in the position-space representation.
 Here ${\cal K}$ is taken to be identical to the $B$-meson LCDA case, $\Gamma_g$ is a constant, and we used 
\begin{align}
    [{\cal D},\, \ln {\cal K}] = 1\, ,\qquad \qquad [{\cal D}, \,\ln \widehat S_-] = -1\, 
\end{align}
to construct the above ansatz. (Note that $\widehat S_-$ receives no quantum corrections and hence its canonical form holds to all orders.) This ansatz also complies with the fact that the $\ln(\widehat S_-/\mu)$ contribution is exclusively generated by the adjoint Wilson lines interacting with the soft quarks, giving rise to the adjoint cusp anomalous dimension. In addition, Eq.~\eqref{eq:gKernel-ansatz} also satisfies the equation
\begin{align}
    s\,[\widehat S_-, \gamma_g(\alpha_s^*;s,\mu)] = -\Gamma_{\rm cusp}^{\rm F}(\alpha_s^*)+\frac12\Gamma_{\rm cusp}^{\rm A}(\alpha_s^*)\, ,
\end{align}
the significance of which calls for further investigation. We did not find an exact way to extract the $\mathcal{O}(\alpha_s^2)$ correction to $\Gamma_g$ from known results of anomalous dimensions of local operators. However, it is reasonable to expect that,
\begin{align}\label{eq:Gamma_g}
    2\Gamma_g = 2\Gamma_\gamma - \Gamma_\wedge + \Delta
\end{align}
with $2\Gamma_g$ being the constant part of $\gamma_g(s,t)$ in the conformal representation (see~\eqref{eq:AM-s;t} and~\eqref{eq:gKernel-ansatz}). 
$\Gamma_\wedge$ denotes the non-cusp anomalous dimension of the vacuum expectation value of the subtraction operator~\eqref{eq:subop} which can be found in~\cite{Erdogan:2011yc,Falcioni:2019nxk},
\begin{align}
    \Gamma_\wedge &= \left(\frac{\alpha_s}{4\pi}\right)^2 C_A\left[C_A\left(\frac{808}{27}-\frac{22}3\zeta_2-4\zeta_3\right)+T_F n_f \left(-\frac{224}{27}+\frac83\zeta_2\right) \right] + {\cal O}(\alpha_s^3)\, .
\end{align}
Interestingly, it differs from the so-called eikonal function $f_{\rm eik}$~\cite{Falcioni:2019nxk} extractable from the Dokshitzer–Gribov–Lipatov–Altarelli–Parisi (DGLAP) kernel, the anomalous dimension $\gamma_s^{\rm DY}$ for the Drell-Yan soft function~\cite{Becher:2007ty}, and the soft anomalous dimension near threshold \cite{Beneke:2009rj} only by a $\zeta_3$ term.
It is also worth noting that all these anomalous dimensions exhibit Casimir scaling at the two-loop order. $\Delta$ in~\eqref{eq:Gamma_g} accounts for the mismatch  arising from contributions beyond $\Gamma_\gamma$ and $\Gamma_\wedge$ that start to appear from the two-loop order. By imposing the consistency relation required by the factorization theorem and using the two-loop hard and jet ($J_g$) function available at the two-loop order in \cite{Liu:2021mac}, we can fix $\Delta$ in~\eqref{eq:Gamma_g} to be 
\begin{equation}
	\label{eq:mismatch}
	\Delta = 
 \left(\frac{\alpha_s}{4\pi}\right)^2 8C_A(C_A+4C_F\big)\,\zeta_3+ {\cal O}(\alpha_s^3)\,,
\end{equation}
which again contains only a $\zeta_3$ term. 

\section{Conclusions}
\label{sect:conclusions}

We established a framework to obtain the renormalization of the soft-quark functions that appear in the context of Higgs form factors 
generated by light-quark loops by direct computation from their operator definitions. While the renormalization-group equations for these 
functions were already known under the assumption that the factorization formula for the physical form factors holds \cite{Liu:2020eqe,Liu:2022ajh}, the direct 
computation provides an independent confirmation of this assumption. 
For the $\gamma\gamma\to h$ case, the position-space renormalization 
technique employed in the present work provides a significant simplification relative to the previous direct computation \cite{Bodwin:2021cpx}. In 
case of $gg\to h$, it reveals that the operator definition that follows 
from the naive application  of SCET factorization  must be amended by a 
subtraction, related to the in-coming colour charges, which turns out to be conceptually identical to the subtraction necessary for the 
factorization of electromagnetic corrections to the form factor in $B_s\to \mu^+\mu^-$ decay \cite{Beneke:2019slt}. Perhaps most importantly, the 
methods developed here open a path to compute the renormalization 
of other soft functions that have recently appeared in the context of hard-scattering factorization beyond the leading power.

For the case of the soft operator $O_\gamma$ relevant to $\gamma\gamma\to h$, we established an all-order relation of its anomalous dimension to the 
one that appears for the leading-twist $B$-meson light-cone distribution amplitude. 
The present work also explores for the first time the consequences of 
conformal symmetry, which  has been successfully applied to the 
computation of the two-loop renormalization of the leading-twist heavy-light string operator \cite{Braun:2019wyx}, for non-string light-cone operators such as $O_\gamma$ and $O_g$ with the latter containing additional complications due to semi-infinite Wilson lines. Thus, it would be highly beneficial from both the theoretical and phenomenological perspective to have a direct two-loop calculation for $\gamma_g$ to verify the conjecture~\eqref{eq:gKernel-ansatz} and track down the origin of the discrepancy $\Delta$ in~\eqref{eq:mismatch}. Further 
investigations and applications of collinear conformal and position-space 
techniques may expose ``hidden relations" that explain why various 
anomalous dimensions differ at the two-loop order only by 
$\zeta_3$ terms. 

\subsection*{Acknowledgements}

We thank A. Belitsky, P. B\"oer, V. Braun, A. Manashov, E. S\"underhauf, A. Vladimirov 
and J.~Wang for discussions. This research was supported in part by the Deutsche
Forschungsgemeinschaft (DFG, German Research Foundation) through
the Sino-German Collaborative Research Center TRR110 ``Symmetries
and the Emergence of Structure in QCD'' 
(DFG Project-ID 196253076, NSFC Grant No. 12070131001, - TRR 110)
and by the Excellence Cluster ORIGINS funded by the Deutsche Forschungsgemeinschaft (DFG, German Research Foundation) under Grant No.~EXC - 2094 - 390783311. 

\appendix
\section*{Appendix}
\addcontentsline{toc}{section}{Appendices}

\renewcommand{\theequation}{\Alph{section}.\arabic{equation}}
\renewcommand{\thetable}{\Alph{table}}
\setcounter{section}{0}
\setcounter{table}{0}

\section{Fourier transform}
\label{app:FT}

In this appendix, we provide some details regarding the Fourier transform~\eqref{eq:softdef} and the application of~\eqref{eq:FT1} and~\eqref{eq:FT2}. 
We start by splitting the logarithmic term in~\eqref{eq:Ogammare} and~\eqref{eq:1Lgamma-pos} into two pieces,
\begin{align}\label{eq:ln=2ln}
    \ln \left(st\mu^2\e^{2\gamma_E}\right) &= \ln(is\mu\e^{\gamma_E} ) + \ln(-it\mu \e^{\gamma_E})\,,
\end{align}
in accordance with~\eqref{eq:FT1} and the $t\to t+i0^+, s\to s-i0^+$ prescriptions are always understood. The Fourier transform over $s$ then reads
\begin{align}
    O_{L_1}(\ell_+,\ell_-) &= \int^\infty_0 \df \ell_+' \df \ell_-'\int_{-\infty}^\infty \df s\,\int^{\infty}_{-\infty}\df t\, \e^{ is(\ell_+-\ell_+')-it(\ell_--\ell_-')}\ln(is\mu\e^{\gamma_E})  O_{\gamma} (\ell_+',\ell_-') \notag\\
    &\hspace*{-20mm}= -\int^\infty_0 \df \ell_+' \df \ell_-' \left[\ln\left(\frac{\ell^+}{\mu}\right)\delta(\ell_+'-\ell_+)+\ell_+\left[\frac{\theta(\ell_+-\ell_+')}{\ell_+(\ell_+-\ell_+')}\right]_+\right]\delta(\ell_--\ell_-')  O_{\gamma} (\ell_+',\ell_-') \, ,
    \end{align}
where we applied~\eqref{eq:FT1}. We next change variables to $\ell_-=w/\ell_+$, $\ell_-'=w/\ell_+'$, leading to
\begin{align}\label{eq:OL1}
        O_{L_1}(\ell_+,\ell_-) &= -\int^\infty_0 \df w'\, \frac{\df \ell_+'}{\ell_+'} \left[\ln\left(\frac{\ell_+}{\mu}\right)\delta(\ell_+'-\ell_+)+\ell_+\left[\frac{\theta(\ell_+-\ell_+')}{\ell_+(\ell_+-\ell_+')}\right]_+\right]  \notag\\
        &\hspace*{10mm}\times \delta\left(\frac{w}{\ell_+}-\frac{w'}{\ell_+'}\right) O_{\gamma} (w') \notag\\[0.2cm]
        &=- \int^\infty_0 \df\omega'\left[\ln\left(\frac{\ell_+}{\mu}\right)\delta(w-w') + w\left[\frac{\theta(w-w')}{w(w-w')}\right]_+\right] O_\gamma (\omega')\, .
\end{align}
We used the ``initial condition" $O_\gamma(\ell_+',\ell_-')=O_\gamma(w')$ thanks to Lorentz  invariance, namely, $O_\gamma$ in momentum space only depends on the scalar product $\ell'^2=\ell_+'\ell_-'$ with $\ell'^\mu = \ell_+' n_-^\mu/2 + \ell_-' n_+^\mu/2$. The Fourier transform over $t$ in~\eqref{eq:ln=2ln} can be carried out straightforwardly as a complex conjugate to~\eqref{eq:OL1} yielding,
\begin{align}
    O_{L_2}(\ell_+,\ell_-) &=- \int^\infty_0 \df\omega'\left[\ln\left(\frac{\ell_-}{\mu}\right)\delta(w-w') + w\left[\frac{\theta(w-w')}{w(w-w')}\right]_+\right] O_\gamma (\omega')\, .
\end{align}
Thus in total, the logarithmic contribution $\ln \left(st\mu^2\e^{2\gamma_E}\right)$ in momentum space reads
\begin{align}
    O_L(\ell_+,\ell_-)= - \int^\infty_0 \df\omega'\left[\ln\left(\frac{w}{\mu^2}\right)\delta(w-w') + 2w\left[\frac{\theta(w-w')}{w(w-w')}\right]_+\right] O_\gamma (\omega')\, .
\end{align}
Following the same procedure and making use of the identity~\eqref{eq:FT2}, the Fourier transform for the remaining position-space renormalization factor in~\eqref{eq:1Lgamma-pos} can be worked out.

\section{Collinear function with an off-shell external gluon}\label{app:jet}

To extract the collinear function, we sandwich the matching equation~\eqref{eq:Jetdef} between a soft quark state with momentum $p_s$ and a slightly off-shell gluon. We then project out the soft quark and the soft Wilson line leaving a collinear vacuum to 
gluon matrix element of collinear fields.  This turns \eqref{eq:Jetdef} into
\begin{eqnarray}
    \label{eq:Jetdefsimple}
     &&\,i\int \df^4 z \,e^{ip_s^+\cdot z}\,\langle 0| \textbf{T}\left\{\left[ \slashed{\mathcal A}_{\perp,n_-}^{(0)} W_{n_-}^{(0)^\dagger}\xi^{(0)}_{n_-}\right](z), \left[\bar{\xi}^{(0)}_{n_-} W_{n_-}^{(0)}\right](\lambda n_+)\right\}|g(k, a)\rangle \\
    &&= \, 2\pi \int\df s\int \df \left(\frac{n_+\cdot z}{2}\right)\,e^{ip_s\cdot z_-+iu n_+\cdot k}\,\tilde{J}_g\left(\lambda,s, \frac{n_+\cdot z}{2}\right) \langle 0| \slashed{\mathcal A}^{(0)}_{\perp, n_-}(sn_+) |g(k,a)\rangle\, .\notag
\end{eqnarray}
The soft quark momentum enters the collinear function only with its $n_+$-component $p_{s^+}^\mu = \frac12 (n_-\cdot p_s) n_+^\mu$ due to multipole expansion. The off-shell momentum-space collinear function then depends on $p^2=(p_s^++k)^2=2p_s^+\cdot k+\mathcal{O}(k^2)$ and  $k^2$.  

\begin{figure}[t]
    \centering
    \includegraphics[width=0.9\textwidth]{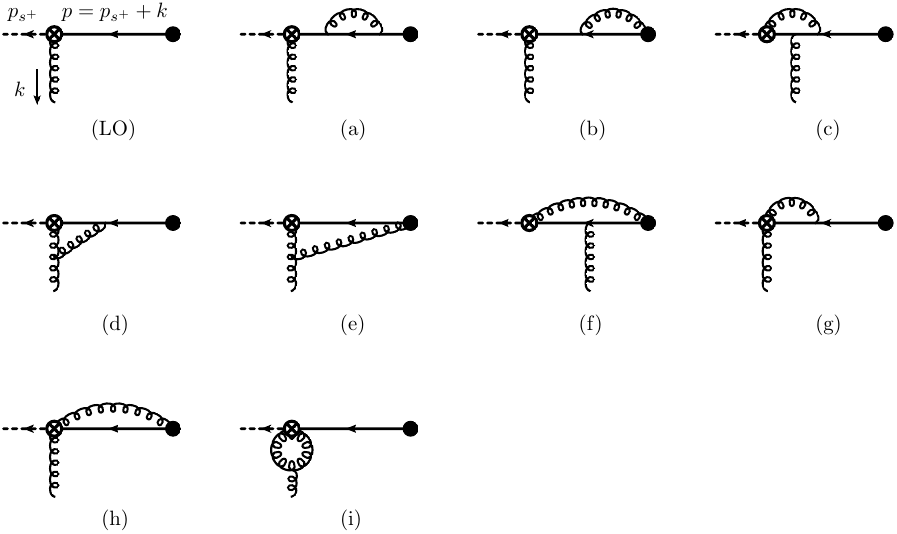}
    \caption{Feynman diagrams contributing to the left-hand side of~\eqref{eq:Jetdefsimple} to $\mathcal{O}(\alpha_s)$ in $R_\xi$ gauge (with and without the background field formalism) using the original subleading power SCET Lagrangian ${\cal L}^{(1)}_{\xi q} (z)= \bar q_s(z_-) W_{n_-}^\dagger(z) i \slashed{D}_{n_-}^\perp \xi_{n_-}(z)$ insertion. The dashed line represents the (decoupled) soft quark field, the symbols ``$\otimes$" and ``{\Large\textbullet}" denote the ${\cal L}_{\xi q}^{(1)}$ insertion and  $[\bar\xi_{n_-}W_{n_-}](0)$, respectively. Note the presence of a symmetry factor $1/2$ of diagram (i). The contribution from the one-loop gluon vacuum polarization cancels between the left- and right-hand side of~\eqref{eq:Jetdefsimple} and is therefore not shown.}
    \label{fig:Jet-qDxi-L}
\end{figure}

Implementing the relevant Feynman rules listed in the appendix of~\cite{Beneke:2018rbh} for the subleading power SCET Lagrangian, we find that the diagrams in Fig.~\ref{fig:Jet-qDxi-L} contribute to the off-shell jet function in $\overline{\rm MS}$-scheme as
\begin{align}\label{eq:J-ls-qDxi}
       D_{g,({\rm LO})}^{\mathrm{OfS}, (0)} &=1\, ,\notag\\
       D_{g,({\rm a})}^{\mathrm{OfS}, (1)} &= -\frac{\alpha_s C_F \xi}{4\pi}\e^{\varepsilon\gamma_E}\left(\frac{-p^2}{\mu^2}\right)^{-\varepsilon}\frac{\Gamma(2-\varepsilon)\Gamma(1-\varepsilon)\Gamma(\varepsilon)}{\Gamma(2-2\varepsilon)}\, ,\notag\\
       D_{g,(\rm b)}^{\mathrm{OfS}, (1)} &= -\frac{\alpha_s C_F}{4\pi}\e^{\varepsilon\gamma_E}\left(\frac{-p^2}{\mu^2}\right)^{-\varepsilon}\left(2-(1-\xi)\varepsilon\right)\frac{\Gamma(2-\varepsilon)\Gamma(-\varepsilon)\Gamma(\varepsilon)}{\Gamma(2-2\varepsilon)}\, ,\notag\\
       D_{g,(\rm c)}^{\mathrm{OfS}, (1)} &=\frac{\alpha_s}{4\pi}\left(C_F-\frac{C_A}2\right)\e^{\varepsilon\gamma_E}\left(\frac{-p^2}{\mu^2}\right)^{-\varepsilon}\left(2-\varepsilon+2\varepsilon^2\right)\frac{\Gamma(-\varepsilon)\Gamma(1-\varepsilon)\Gamma(\varepsilon)}{\Gamma(2-2\varepsilon)}\, ,\notag\\
       D_{g,(\rm d)}^{\mathrm{OfS}, (1)} &=\frac{\alpha_s C_A}{4\pi}\e^{\varepsilon\gamma_E}\left(\frac{-p^2}{\mu^2}\right)^{-\varepsilon}\left(2-(2+\xi)\varepsilon-(1-\xi)\varepsilon^2\right)\frac{\Gamma(-\varepsilon)\Gamma(1-\varepsilon)\Gamma(\varepsilon)}{2\Gamma(2-2\varepsilon)}\, ,\notag\\
       D_{g,(\rm e)}^{\mathrm{OfS}, (1)} &=\frac{\alpha_sC_A}{4\pi}\e^{\varepsilon\gamma_E}\left[\left(\frac{-p^2}{\mu^2}\right)^{-\varepsilon}\left(2-(5-\xi)\varepsilon+(1-\xi)\varepsilon^2\right)\right. \notag\\
            &\quad\left.-\left(-\frac{k^2}{\mu^2}\right)^{-\varepsilon}\left(2-(5-\xi)\varepsilon+2(1-\xi)\varepsilon^2\right)\right]\frac{\Gamma(-\varepsilon)\Gamma(1-\varepsilon)\Gamma(\varepsilon)}{2\Gamma(2-2\varepsilon)}\, ,\notag\\
       D_{g,(\rm f)}^{\mathrm{OfS}, (1)} &=0\,, \\[0.1cm]
       D_{g,(\rm g)}^{\mathrm{OfS}, (1)} &= -\frac{\alpha_s}{4\pi}\left(C_F-\frac{C_A}2\right)\e^{\varepsilon\gamma_E}\left(\frac{-p^2}{\mu^2}\right)^{-\varepsilon}\left(2-(1-\xi)\varepsilon\right)\frac{\Gamma(2-\varepsilon)\Gamma(-\varepsilon)\Gamma(\varepsilon)}{\Gamma(2-2\varepsilon)}\, ,\notag\\
       D_{g,(\rm h)}^{\mathrm{OfS}, (1)} &=\frac{\alpha_s}{4\pi}\left(C_F-\frac{C_A}2\right)\e^{\varepsilon\gamma_E}\left(\frac{-p^2}{\mu^2}\right)^{-\varepsilon}(1-\xi)\frac{\Gamma(2-\varepsilon)\Gamma(1-\varepsilon)\Gamma(\varepsilon)}{\Gamma(2-2\varepsilon)}\, ,\notag\\
       D_{g,(\rm i)}^{\mathrm{OfS}, (1)} &=-\frac{\alpha_sC_A}{4\pi}\e^{\varepsilon\gamma_E}\left(-\frac{k^2}{\mu^2}\right)^{-\varepsilon}(2-\varepsilon+\xi\varepsilon)(2-3\varepsilon-\xi\varepsilon)\frac{\Gamma(1-\varepsilon)\Gamma(-\varepsilon)\Gamma(\varepsilon)}{4\Gamma(2-2\varepsilon)}\, .\notag
       \end{align}
The calculation is performed in the background field $R_\xi$ gauge~\cite{Abbott:1981ke}. Alternatively, conducting the calculation in $R_\xi$ gauge without employing the background field method leads to identical result. The virtuality of the external gluon serves as an infrared (IR) regulator, implying $k^2\ll p^2$. Consequently, in~\eqref{eq:J-ls-qDxi}, we retain only the non-vanishing terms in the $k^2\to 0$ limit.

\begin{figure}[t]
    \centering
     \includegraphics[width=0.2\textwidth]{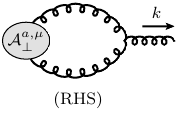}
    \caption{Feynman diagram responsible for the right-hand side (RHS) contribution of the matching equation~\eqref{eq:Jetdefsimple}, i.e., the matrix element of the gauge-invariant building block ${\cal A}$~\eqref{eq:OSRHS}. Similar to diagram (i) in Fig.~\ref{fig:Jet-qDxi-L}, it also comes with a symmetry factor $1/2$. The vacuum polarization diagram cancels between the left- and right-hand side of the matching equation and, hence, can be dropped.}
    \label{fig:Jet-qDxi-R}
\end{figure}

The right-hand side of the matching equation~\eqref{eq:Jetdefsimple} receives quantum corrections when the external gluon is off-shell. At the one-loop level, they are generated by the diagram depicted in Fig.~\ref{fig:Jet-qDxi-R}. Its contribution to the off-shell collinear function at NLO reads,
\begin{align}\label{eq:J-RHS}
    \mathfrak{A}_g^{-, (\xi)} &= \frac{\alpha_s C_A}{4\pi}\left(-\frac{\mu^2\e^{\gamma_E}}{k^2}\right)^{\varepsilon}\left(8-2(9-\xi)\varepsilon+(1-\xi)(7+\xi)\varepsilon^2\right)\frac{\Gamma^2(-\varepsilon)\Gamma(1+\varepsilon)}{4\Gamma(2-2\varepsilon)}\,.
\end{align}
Here, we have neglected the contribution from the gluon vacuum polarization as it cancels between the left- and right-hand side of~\eqref{eq:Jetdefsimple}. From ~\eqref{eq:J-ls-qDxi} and~\eqref{eq:J-RHS}, we find the off-shell collinear function to ${\cal O}(\alpha_s)$ to be,
\begin{align}
    J_{g}^{\mathrm{OfS}, (0)} &= D_{g,({\rm LO})}^{\mathrm{OfS}, (0)} =1\, ,\notag\\
    J_{g}^{\mathrm{OfS}, (1)} &=  D_{g,{\rm (a)}}^{\mathrm{OfS}, (1)} +\cdots + D_{g,{\rm (i)}}^{\mathrm{OfS}, (1)}- \mathfrak{A}_g^{-, (\xi)} \notag\\
    &= \frac{\alpha_s}{4\pi}(C_A-C_F)\left(-\frac{2}{\varepsilon^2}+\frac{2\ln(-p^2/\mu^2)}{\varepsilon}\right) + {\cal O}(\varepsilon^0) \notag\\
    &\equiv J_{g,1/\varepsilon}^{\mathrm{OfS}, (1)} + {\cal O}(\varepsilon^0) = J_{g,1/\varepsilon}^{\mathrm{OS}, (1)} + {\cal O}(\varepsilon^0) \, ,\label{eq:JofS-final}
\end{align}
where we performed the $\varepsilon$-expansion to ${\cal O}(1/\varepsilon)$, since we are only interested in the UV renormalization of the function. Remarkably, the $\varepsilon$-divergence of the off-shell collinear function coincides with the on-shell one. It is also important to note that while both sides of the matching equation~\eqref{eq:Jetdefsimple} exhibit gauge-dependence in the single-pole term $1/\varepsilon$, the off-shell collinear function defined as a matching coefficient remains \textit{gauge-independent}. 

\begin{figure}[t]
    \centering
    \includegraphics[width=0.85\textwidth]{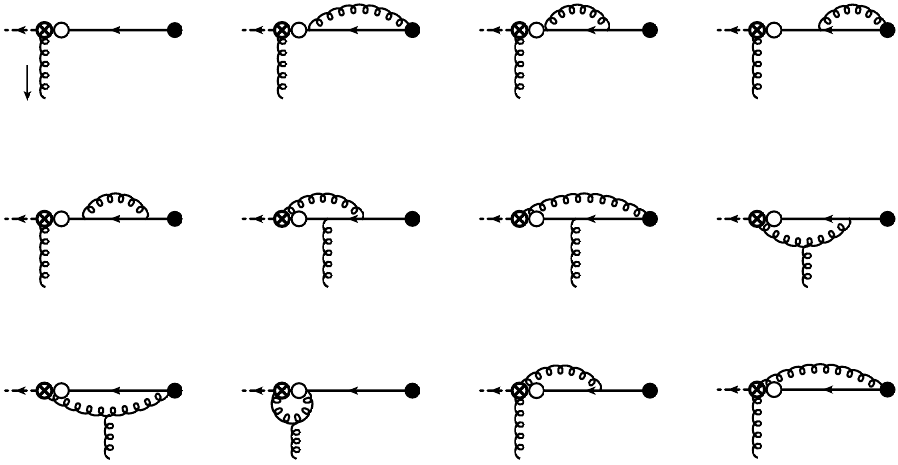}
    \caption{Feynman diagrams contributing to the off-shell collinear function $J_g$ at the one-loop level using the gauge-invariant SCET building blocks ${\mathcal A}_{n_-}^{\perp \mu}$ and $\chi_{n_-}$. The symbols ``$\otimes$", ``{\Large$\circ$}" and ``{\Large\textbullet}" denote the fields $\slashed{\mathcal A}_{n_-}^\perp(z)$, $\chi_{n_-}(z)$, and $\chi_{n_-}(0)$, respectively. Diagrams only involving ${\mathcal A}_{n_-}^{\perp\mu}$ (i.e.~Fig.~\ref{fig:Jet-qDxi-R} and the one-loop gluon vacuum polarization correction to the first diagram in Fig.~\ref{fig:Jet-qDxi-L}) are neglected since their contributions are canceled between the two sides of~\eqref{eq:Jetdefsimple}.}
    \label{fig:jet-qAchi}
\end{figure}

Equivalently, the collinear function can be computed in terms of the gauge-invariant collinear building blocks in SCET, in which case the subleading SCET Lagrangian takes the form ${\cal L}^{(1)}_{q{\cal A}\chi} = \bar q_s(z_-)\slashed{\mathcal A}^\perp_{n_-}(z)\chi_{n_-}(z)$ with ${\mathcal A}^{\perp\mu}_{n_-} (z) = [W_{n_-}^\dagger i D_\perp^\mu W_{n_-}](z)$ and $\chi_{n_-}(z) = [W_{n_-}^\dagger\xi_{n_-}](z)$. The relevant Feynman diagrams in this approach are collected in Fig.~\ref{fig:jet-qAchi}. Computing the jet function using the SCET building blocks provides a valuable cross-check of the result~\eqref{eq:JofS-final}. In particular, in this case, there is no need to calculate corrections to the right-hand side of the matching equation~\eqref{eq:Jetdefsimple}, as they are entirely canceled by the corresponding one-loop diagrams on the left-hand side, where only $\slashed{\mathcal A}_{n-}^\perp$ contributes to the interaction (not shown in~Fig.~\ref{fig:jet-qAchi}). Performing the one-loop calculation in background-field $R_\xi$ gauge and summing up all contributing diagrams confirms the expression~\eqref{eq:JofS-final} for  $J_{g}^{\mathrm{OfS}, (1)} $ to all orders in $\varepsilon$.

Furthermore, we also computed the off-shell collinear function in the light-cone gauge and again found complete agreement with~\eqref{eq:JofS-final}. The main advantage of adopting the light-cone gauge is that the number of diagrams is smaller. 

\section{Solving RGE using conformal symmetry}
\label{app:conformal}

In this appendix, we investigate the application of conformal symmetry to solving the RGE for $O_\gamma$ and $O_g$ inspired by the study of $B$-LCDA evolution~\cite{Bell:2013tfa,Braun:2014owa,Braun:2015pha,Braun:2017liq,Braun:2018fiz}. 
First, we note that in terms of the collinear conformal generators in \eqref{eq:generator} and~\eqref{eq:generatorT}, the \textit{position-space} renormalization factor 
for $O_g$ in the $\overline{\rm MS}$-scheme can be rewritten compactly as
\begin{align}
	\label{eq:ZSx}
		Z_g =& 1 + \frac{\alpha_s(\mu)}{4\pi}\Bigg\{\frac{2(C_F-C_A)}{\varepsilon^2}+\frac{2C_F-C_A}{\varepsilon}\Big[\ln\left(\mu^2\widehat{S}^+\widehat{T}^+\right)-2+4\gamma_E\Big]\notag\\
		&+\frac{C_A}{\varepsilon}\left[\ln\frac{\widehat{S}_-\widehat{T}_-}{\mu^2}-2\right]+\frac{C_F}{\varepsilon}\Bigg\}+\mathcal{O}(\alpha_s^2)\,.
\end{align}
The corresponding one-loop anomalous dimension reads, 
\begin{eqnarray}
		\gamma_g &=&  \frac{\alpha_s(\mu)}{4\pi}\Bigg[4\left(C_F-\frac{C_A}{2}\right)\Big(\ln\left(\mu^2 \widehat{S}_+ \widehat{T}_+\right)+4\gamma_E-2\Big)+2C_A\left(\ln\frac{\widehat{S}_-\widehat{T}_-}{\mu^2}-2\right)\notag\\
		&&\hspace*{0mm}+\,2C_F\Bigg] 
		\equiv  \frac{\alpha_s(\mu)}{4\pi}\widehat{H}^{(1)}\, .
\label{eq:ADx}\end{eqnarray}
The position-space RGE is simply given by 
\begin{equation}
	\label{eq:RGEx}
	\left[\frac{\df}{\df\ln\mu}+\gamma_g\right] O_g(s,t) =0 \,.
\end{equation}
The non-local information of $\gamma_g$ is hidden in the conformal operators. Since the two sets of generators $\{\widehat{S}_+, \widehat{S}_-, \widehat{S}_0\}$ and $\{\widehat{T}_+, \widehat{T}_-, \widehat{T}_0\}$ commute with each other, the one-loop kernel can be disentangled as $\widehat{H}^{(1)} = \widehat{H}^{(1)}_s+\widehat{H}^{(1)}_t$, with
\begin{equation}
	\label{eq:ADxkernel}
		\widehat{H}^{(1)}_s= 
		4\left(C_F-\frac{C_A}{2}\right)\Big[\ln\big(\mu\widehat{S}_+\big)+2\gamma_E-1\Big]+2C_A\left[\ln \frac{\widehat{S}_-}{\mu}-1\right]+C_F\, ,
\end{equation}
similarly for $\widehat{H}^{(1)}_t$ with 
$\widehat S_\pm \to \widehat T_\pm$. Consequently, the properties of the full one-loop kernel $\widehat H^{(1)}$ are completely determined by $\widehat{H}^{(1)}_s$. Before delving into the full solution of \eqref{eq:ADxkernel}, we take a look at three degenerate cases. 
\begin{enumerate}
	\item {\bf Abelian limit: $C_A=0$}
 
	In this case, $\ln(\widehat{S}_-/\mu)$ is absent, which aligns with the evolution kernel of the leading-twist $B$-LCDA and $O_\gamma$. Therefore, the eigenfunctions of $\widehat{H}^{(1)}_s$ are completely determined by the eigenequation of $\widehat{S}_+$~\cite{Braun:2014owa,Liu:2020eqe}, since the underlying system has only one degree of freedom~(one variable $s$ for $\widehat H^{(1)}_s$). The latter is readily solved by
\begin{equation}
	\label{eq:eigenfuncgamma}
	Q_{s}(z)=-\frac{1}{z^{2}} \mathrm{e}^{-s / z}, \quad \left[\widehat{S}_{+} Q_{s}\right](z) = s Q_{s}(z)\, .
\end{equation}
    \item {\bf Large $N_c$ limit: $2C_F-C_A=0$}
    
    In this limit, only the second term  proportional to $C_A$ in~\eqref{eq:ADxkernel} survives and the eigenfunction becomes the exponential function ${\rm e}^{-s z}$. 
    \item {\bf Supersymmetric limit: $C_F=C_A$}
     
    By trading $\widehat{S}_+$ and $\widehat{S}_-$ in terms of the Euler operator $\theta_s$ \cite{Braun:2014owa}, the kernel simplifies to
    \begin{align}
    	\label{eq:SUSYlimit}
    		\widehat{H}_s^{(1)}\longrightarrow &~~  
    		2C_A\left[\ln\left(\mu\widehat{S}_+\right)+\ln\frac{\widehat{S}_-}{\mu}+2\gamma_E-\frac{3}{2}\right] \notag\\
    		&= 2C_A\left[\psi_0(\theta_s+2)+\psi_0(-\theta_s)-2\psi_0(1)-\frac{3}{2}\right]\, ,
    	\end{align}
    where $\psi_0(z)$ denotes the polygamma function of the zeroth order. 
    The eigenfunctions of $\widehat{H}_s^{(1)}$ are determined by solving the eigenequation of $\theta_s$ with power-law solutions, expressed as $\theta_s\,s^\lambda=\lambda\,s^\lambda$. This scenario makes no distinction between the fundamental and adjoint Wilson lines because all fields live in the adjoint representation of the gauge group. 
\end{enumerate}

\noindent We now turn to the general case \eqref{eq:ADxkernel}. We start from the representation 
\begin{equation}
		\widehat{H}^{(1)}_s= 
		4\left(C_F-\frac{C_A}{2}\right)\Big[\psi_0(\theta_s+1)+\ln(i \mu s)-\psi_0(1)-1\Big]+2C_A\left[\ln \frac{\widehat{S}_-}{\mu}-1\right]+C_F\, ,
\end{equation}
 for $\widehat{H}^{(1)}_s$. Noting that~\cite{Belitsky:2014rba,Braun:2014owa}
\begin{align}
\ln(i\partial_z)  = \psi_0(-\theta_z)-\ln(iz)\, ,
\end{align}
allows us to rewrite ${\widehat H}_S^{(1)}$ as,
\begin{align}
{\widehat H}_s^{(1)} &=4\left(C_F-\frac{C_A}{2}\right)\left(\psi_0(1)-\psi_0(\theta_s+2)\right)-2C_A\psi_0(-\theta_s)-4(C_F-C_A)\ln(\mu s)\notag\\
&+3C_F.
\end{align}
Subsequently, the eigenfunction $\phi_\lambda(s)$ of the evolution kernel ${\widehat H}_s$ can be found in Mellin space~\cite{Braun:2019zhp},
\begin{align}
\phi_\lambda(s) &= \int^{c_0+i\infty}_{c_0-i\infty}\frac{\df j}{2\pi i} \, (\mu_0 s)^{-j}\,\widetilde\phi_\lambda(j)\, ,\label{eq:Mellin}
\end{align}
where $\mu_0$ is an arbitrary dimensionful constant making $(\mu_0 s)$ dimensionless. $c_0$ is a generic non-integer number to avoid potential poles from $\Gamma$ functions. The eigenvalue equation
\begin{align}
 \widehat H_s\phi_\lambda(s) = {\mathbb E}_\lambda \phi_ \lambda (s)
\end{align}
becomes a first-order differential equation for $\widetilde\phi_\lambda(j)$, which reads
\begin{align}
&\left[2(C_F-C_A)\left(\frac{\df}{\df j}+\ln\frac{\mu}{\mu_0}\right)  + 2\left(C_F-\frac{C_A}{2}\right)\left(\psi_0(2-j)-\psi_0(1)-1\right)+C_A\psi_0(j)\right.\notag\\
&\qquad\left.+\frac{C_F}{2}-C_A + \frac{{\mathbb E}_\lambda}{2a_s} \right]\widetilde\phi_\lambda(j) =0\, ,
\end{align}
with the short-hand notation $a_s=\alpha_s/(4\pi)$. The above equation is readily solved by
\begin{align}
\widetilde\phi_\lambda(j) = &\exp\left[-\left(\frac{C_F-2C_A+2\left(2C_F-C_A\right)(\gamma_E-1)}{4(C_F-C_A)} 
\right)j
\right]\notag\\[0.1cm]
&\times\left(\Gamma(2-j)\right)^{\frac{2C_F-C_A}{2C_F-2C_A}}\left(\Gamma(j)\right)^{\frac{C_A}{2C_A-2C_F}}\lambda^j 
\, ,
\end{align}
with eigenvalue ${\mathbb E}_\lambda=4a_s(C_A-C_F)\ln(\mu\lambda/\mu_0)$. Expanding a generic function $\widetilde\phi(j,\mu)$ in the eigenfunction basis $\widetilde \phi_\lambda(j)$ via the inverse Mellin transform, 
\begin{align}\label{eq:Mellin-evo}
\widetilde\phi(j,\mu) & = \int^\infty_0\df\lambda\, \eta(\lambda,\mu)\, \widetilde \phi_\lambda(j)\,,
\end{align}
simplifies the RGE to
\begin{align}
\left[\frac{\df}{\df\ln\mu} + \frac{\alpha_s(\mu)}{4\pi}(4C_A-4C_F)\ln\frac{\lambda\mu}{\mu_0} \right]\eta(\lambda,\mu) = 0\, .
\end{align}
This equation can be solved in the standard way, yielding
\begin{align}
\eta(\lambda,\mu)&=\exp\left[-4\int^\mu_{\mu_0}\frac{\df\tau}{\tau} \frac{\alpha_s(\tau)}{4\pi}(C_A-C_F)\ln\left(\frac{\tau\lambda}{\mu_0}\right)\right] \eta(\lambda,\mu_0)\, .
\end{align}
The final form of the eigenfunction for $\widehat H^{(1)}$ is simply the product $\phi_{\lambda_1}(s,\mu)\phi_{\lambda_2}(t,\mu)$. Compared to other techniques, the conformal symmetry-inspired Mellin-space approach to solving the RGE is systematic and well-suited for accommodating higher-order corrections to the anomalous dimensions~\cite{Braun:2019zhp}. The analytic solution to the one-loop RGE in momentum space can be derived by applying the Fourier and Mellin transform to~\eqref{eq:Mellin-evo} using~\eqref{eq:Mellin}. We  refer~\cite{Liu:2022ajh} for the explicit expression.

\bibliographystyle{JHEP}
\bibliography{paper.bib}

\providecommand{\href}[2]{#2}\begingroup\raggedright\begin{thebibliography}{10}

\bibitem{Korchemsky:1993uz}
G.~P. Korchemsky and G.~Marchesini, \emph{{Resummation of large infrared
  corrections using Wilson loops}},
  \href{https://doi.org/10.1016/0370-2693(93)90015-A}{\emph{Phys. Lett. B}
  {\bfseries 313} (1993) 433--440}.

\bibitem{Bauer:2000yr}
C.~W. Bauer, S.~Fleming, D.~Pirjol and I.~W. Stewart, \emph{{An Effective field
  theory for collinear and soft gluons: Heavy to light decays}},
  \href{https://doi.org/10.1103/PhysRevD.63.114020}{\emph{Phys. Rev. D}
  {\bfseries 63} (2001) 114020},
  [\href{https://arxiv.org/abs/hep-ph/0011336}{{\ttfamily hep-ph/0011336}}].

\bibitem{Beneke:2002ph}
M.~Beneke, A.~P. Chapovsky, M.~Diehl and T.~Feldmann, \emph{{Soft collinear
  effective theory and heavy to light currents beyond leading power}},
  \href{https://doi.org/10.1016/S0550-3213(02)00687-9}{\emph{Nucl. Phys. B}
  {\bfseries 643} (2002) 431--476},
  [\href{https://arxiv.org/abs/hep-ph/0206152}{{\ttfamily hep-ph/0206152}}].

\bibitem{Beneke:2002ni}
M.~Beneke and T.~Feldmann, \emph{{Multipole expanded soft collinear effective
  theory with nonAbelian gauge symmetry}},
  \href{https://doi.org/10.1016/S0370-2693(02)03204-5}{\emph{Phys. Lett. B}
  {\bfseries 553} (2003) 267--276},
  [\href{https://arxiv.org/abs/hep-ph/0211358}{{\ttfamily hep-ph/0211358}}].

\bibitem{Beneke:2004in}
M.~Beneke, F.~Campanario, T.~Mannel and B.~D. Pecjak, \emph{{Power corrections
  to anti-B ---\ensuremath{>} X(u) l anti-nu (X(s) gamma) decay spectra in the
  'shape-function' region}},
  \href{https://doi.org/10.1088/1126-6708/2005/06/071}{\emph{JHEP} {\bfseries
  06} (2005) 071}, [\href{https://arxiv.org/abs/hep-ph/0411395}{{\ttfamily
  hep-ph/0411395}}].

\bibitem{Lee:2004ja}
K.~S.~M. Lee and I.~W. Stewart, \emph{{Factorization for power corrections to B
  ---\ensuremath{>} X(s) gamma and B ---\ensuremath{>} X(u) l anti-nu}},
  \href{https://doi.org/10.1016/j.nuclphysb.2005.05.004}{\emph{Nucl. Phys. B}
  {\bfseries 721} (2005) 325--406},
  [\href{https://arxiv.org/abs/hep-ph/0409045}{{\ttfamily hep-ph/0409045}}].

\bibitem{Bosch:2004cb}
S.~W. Bosch, M.~Neubert and G.~Paz, \emph{{Subleading shape functions in
  inclusive B decays}},
  \href{https://doi.org/10.1088/1126-6708/2004/11/073}{\emph{JHEP} {\bfseries
  11} (2004) 073}, [\href{https://arxiv.org/abs/hep-ph/0409115}{{\ttfamily
  hep-ph/0409115}}].

\bibitem{Moult:2018jjd}
I.~Moult, I.~W. Stewart, G.~Vita and H.~X. Zhu, \emph{{First Subleading Power
  Resummation for Event Shapes}},
  \href{https://doi.org/10.1007/JHEP08(2018)013}{\emph{JHEP} {\bfseries 08}
  (2018) 013}, [\href{https://arxiv.org/abs/1804.04665}{{\ttfamily
  1804.04665}}].

\bibitem{Beneke:2018gvs}
M.~Beneke, A.~Broggio, M.~Garny, S.~Jaskiewicz, R.~Szafron, L.~Vernazza et~al.,
  \emph{{Leading-logarithmic threshold resummation of the Drell-Yan process at
  next-to-leading power}},
  \href{https://doi.org/10.1007/JHEP03(2019)043}{\emph{JHEP} {\bfseries 03}
  (2019) 043}, [\href{https://arxiv.org/abs/1809.10631}{{\ttfamily
  1809.10631}}].

\bibitem{Beneke:2019oqx}
M.~Beneke, A.~Broggio, S.~Jaskiewicz and L.~Vernazza, \emph{{Threshold
  factorization of the Drell-Yan process at next-to-leading power}},
  \href{https://doi.org/10.1007/JHEP07(2020)078}{\emph{JHEP} {\bfseries 07}
  (2020) 078}, [\href{https://arxiv.org/abs/1912.01585}{{\ttfamily
  1912.01585}}].

\bibitem{Broggio:2021fnr}
A.~Broggio, S.~Jaskiewicz and L.~Vernazza, \emph{{Next-to-leading power
  two-loop soft functions for the Drell-Yan process at threshold}},
  \href{https://doi.org/10.1007/JHEP10(2021)061}{\emph{JHEP} {\bfseries 10}
  (2021) 061}, [\href{https://arxiv.org/abs/2107.07353}{{\ttfamily
  2107.07353}}].

\bibitem{Broggio:2023pbu}
A.~Broggio, S.~Jaskiewicz and L.~Vernazza, \emph{{Threshold factorization of
  the Drell-Yan quark-gluon channel and two-loop soft function at
  next-to-leading power}},  \href{https://arxiv.org/abs/2306.06037}{{\ttfamily
  2306.06037}}.

\bibitem{Beneke:2019mua}
M.~Beneke, M.~Garny, S.~Jaskiewicz, R.~Szafron, L.~Vernazza and J.~Wang,
  \emph{{Leading-logarithmic threshold resummation of Higgs production in gluon
  fusion at next-to-leading power}},
  \href{https://doi.org/10.1007/JHEP01(2020)094}{\emph{JHEP} {\bfseries 01}
  (2020) 094}, [\href{https://arxiv.org/abs/1910.12685}{{\ttfamily
  1910.12685}}].

\bibitem{Liu:2019oav}
Z.~L. Liu and M.~Neubert, \emph{{Factorization at subleading power and
  endpoint-divergent convolutions in $h\to\gamma\gamma$ decay}},
  \href{https://doi.org/10.1007/JHEP04(2020)033}{\emph{JHEP} {\bfseries 04}
  (2020) 033}, [\href{https://arxiv.org/abs/1912.08818}{{\ttfamily
  1912.08818}}].

\bibitem{Liu:2020wbn}
Z.~L. Liu, B.~Mecaj, M.~Neubert and X.~Wang, \emph{{Factorization at subleading
  power and endpoint divergences in $h\to\gamma\gamma$ decay. Part II.
  Renormalization and scale evolution}},
  \href{https://doi.org/10.1007/JHEP01(2021)077}{\emph{JHEP} {\bfseries 01}
  (2021) 077}, [\href{https://arxiv.org/abs/2009.06779}{{\ttfamily
  2009.06779}}].

\bibitem{Liu:2022ajh}
Z.~L. Liu, M.~Neubert, M.~Schnubel and X.~Wang, \emph{{Factorization at
  next-to-leading power and endpoint divergences in $gg \to h$ production}},
  \href{https://doi.org/10.1007/JHEP06(2023)183}{\emph{JHEP} {\bfseries 06}
  (2023) 183}, [\href{https://arxiv.org/abs/2212.10447}{{\ttfamily
  2212.10447}}].

\bibitem{Liu:2020eqe}
Z.~L. Liu, B.~Mecaj, M.~Neubert, X.~Wang and S.~Fleming, \emph{{Renormalization
  and Scale Evolution of the Soft-Quark Soft Function}},
  \href{https://doi.org/10.1007/JHEP07(2020)104}{\emph{JHEP} {\bfseries 07}
  (2020) 104}, [\href{https://arxiv.org/abs/2005.03013}{{\ttfamily
  2005.03013}}].

\bibitem{Bodwin:2021cpx}
G.~T. Bodwin, J.-H. Ee, J.~Lee and X.-P. Wang, \emph{{Analyticity,
  renormalization, and evolution of the soft-quark function}},
  \href{https://doi.org/10.1103/PhysRevD.104.016010}{\emph{Phys. Rev. D}
  {\bfseries 104} (2021) 016010},
  [\href{https://arxiv.org/abs/2101.04872}{{\ttfamily 2101.04872}}].

\bibitem{Lange:2003ff}
B.~O. Lange and M.~Neubert, \emph{{Renormalization group evolution of the B
  meson light cone distribution amplitude}},
  \href{https://doi.org/10.1103/PhysRevLett.91.102001}{\emph{Phys. Rev. Lett.}
  {\bfseries 91} (2003) 102001},
  [\href{https://arxiv.org/abs/hep-ph/0303082}{{\ttfamily hep-ph/0303082}}].

\bibitem{Braun:2003wx}
V.~M. Braun, D.~Y. Ivanov and G.~P. Korchemsky, \emph{{The B meson distribution
  amplitude in QCD}},
  \href{https://doi.org/10.1103/PhysRevD.69.034014}{\emph{Phys. Rev. D}
  {\bfseries 69} (2004) 034014},
  [\href{https://arxiv.org/abs/hep-ph/0309330}{{\ttfamily hep-ph/0309330}}].

\bibitem{Beneke:2019slt}
M.~Beneke, C.~Bobeth and R.~Szafron, \emph{{Power-enhanced leading-logarithmic
  QED corrections to $B_q \to \mu^+\mu^-$}},
  \href{https://doi.org/10.1007/JHEP10(2019)232}{\emph{JHEP} {\bfseries 10}
  (2019) 232}, [\href{https://arxiv.org/abs/1908.07011}{{\ttfamily
  1908.07011}}].

\bibitem{Beneke:2022msp}
M.~Beneke, P.~B\"oer, J.-N. Toelstede and K.~K. Vos, \emph{{Light-cone
  distribution amplitudes of heavy mesons with QED effects}},
  \href{https://doi.org/10.1007/JHEP08(2022)020}{\emph{JHEP} {\bfseries 08}
  (2022) 020}, [\href{https://arxiv.org/abs/2204.09091}{{\ttfamily
  2204.09091}}].

\bibitem{Braun:2003rp}
V.~M. Braun, G.~P. Korchemsky and D.~M\"uller, \emph{{The Uses of conformal
  symmetry in QCD}},
  \href{https://doi.org/10.1016/S0146-6410(03)90004-4}{\emph{Prog. Part. Nucl.
  Phys.} {\bfseries 51} (2003) 311--398},
  [\href{https://arxiv.org/abs/hep-ph/0306057}{{\ttfamily hep-ph/0306057}}].

\bibitem{Braun:2013tva}
V.~M. Braun and A.~N. Manashov, \emph{{Evolution equations beyond one loop from
  conformal symmetry}},
  \href{https://doi.org/10.1140/epjc/s10052-013-2544-1}{\emph{Eur. Phys. J. C}
  {\bfseries 73} (2013) 2544},
  [\href{https://arxiv.org/abs/1306.5644}{{\ttfamily 1306.5644}}].

\bibitem{Braun:2016qlg}
V.~M. Braun, A.~N. Manashov, S.~Moch and M.~Strohmaier, \emph{{Two-loop
  conformal generators for leading-twist operators in QCD}},
  \href{https://doi.org/10.1007/JHEP03(2016)142}{\emph{JHEP} {\bfseries 03}
  (2016) 142}, [\href{https://arxiv.org/abs/1601.05937}{{\ttfamily
  1601.05937}}].

\bibitem{Beneke:2020vnb}
M.~Beneke, P.~B\"oer, J.-N. Toelstede and K.~K. Vos, \emph{{QED factorization
  of non-leptonic $B$ decays}},
  \href{https://doi.org/10.1007/JHEP11(2020)081}{\emph{JHEP} {\bfseries 11}
  (2020) 081}, [\href{https://arxiv.org/abs/2008.10615}{{\ttfamily
  2008.10615}}].

\bibitem{Braun:2019wyx}
V.~M. Braun, Y.~Ji and A.~N. Manashov, \emph{{Two-loop evolution equation for
  the B-meson distribution amplitude}},
  \href{https://doi.org/10.3204/PUBDB-2019-02451}{\emph{Phys. Rev. D}
  {\bfseries 100} (2019) 014023},
  [\href{https://arxiv.org/abs/1905.04498}{{\ttfamily 1905.04498}}].

\bibitem{Echevarria:2015byo}
M.~G. Echevarria, I.~Scimemi and A.~Vladimirov, \emph{{Universal transverse
  momentum dependent soft function at NNLO}},
  \href{https://doi.org/10.1103/PhysRevD.93.054004}{\emph{Phys. Rev. D}
  {\bfseries 93} (2016) 054004},
  [\href{https://arxiv.org/abs/1511.05590}{{\ttfamily 1511.05590}}].

\bibitem{Braun:2017liq}
V.~M. Braun, Y.~Ji and A.~N. Manashov, \emph{{Higher-twist B-meson Distribution
  Amplitudes in HQET}},
  \href{https://doi.org/10.1007/JHEP05(2017)022}{\emph{JHEP} {\bfseries 05}
  (2017) 022}, [\href{https://arxiv.org/abs/1703.02446}{{\ttfamily
  1703.02446}}].

\bibitem{Huang:2023jdu}
Y.-K. Huang, Y.~Ji, Y.-L. Shen, C.~Wang, Y.-M. Wang and X.-C. Zhao,
  \emph{{Renormalization-Group Evolution for the Bottom-Meson Soft Function}},
  \href{https://arxiv.org/abs/2312.15439}{{\ttfamily 2312.15439}}.

\bibitem{Abbott:1981ke}
L.~F. Abbott, \emph{{Introduction to the Background Field Method}}, {\emph{Acta
  Phys. Polon. B} {\bfseries 13} (1982) 33}.

\bibitem{Balitsky:1987bk}
I.~I. Balitsky and V.~M. Braun, \emph{{Evolution Equations for QCD String
  Operators}}, \href{https://doi.org/10.1016/0550-3213(89)90168-5}{\emph{Nucl.
  Phys. B} {\bfseries 311} (1989) 541--584}.

\bibitem{Becher:2009cu}
T.~Becher and M.~Neubert, \emph{{Infrared singularities of scattering
  amplitudes in perturbative QCD}},
  \href{https://doi.org/10.1103/PhysRevLett.102.162001}{\emph{Phys. Rev. Lett.}
  {\bfseries 102} (2009) 162001},
  [\href{https://arxiv.org/abs/0901.0722}{{\ttfamily 0901.0722}}].

\bibitem{Beneke:2020ibj}
M.~Beneke, M.~Garny, S.~Jaskiewicz, R.~Szafron, L.~Vernazza and J.~Wang,
  \emph{{Large-x resummation of off-diagonal deep-inelastic parton scattering
  from d-dimensional refactorization}},
  \href{https://doi.org/10.1007/JHEP10(2020)196}{\emph{JHEP} {\bfseries 10}
  (2020) 196}, [\href{https://arxiv.org/abs/2008.04943}{{\ttfamily
  2008.04943}}].

\bibitem{Liu:2021mac}
Z.~L. Liu, M.~Neubert, M.~Schnubel and X.~Wang, \emph{{Radiative quark jet
  function with an external gluon}},
  \href{https://doi.org/10.1007/JHEP02(2022)075}{\emph{JHEP} {\bfseries 02}
  (2022) 075}, [\href{https://arxiv.org/abs/2112.00018}{{\ttfamily
  2112.00018}}].

\bibitem{Belitsky:2014rba}
A.~V. Belitsky, S.~E. Derkachov and A.~N. Manashov, \emph{{Quantum mechanics of
  null polygonal Wilson loops}},
  \href{https://doi.org/10.1016/j.nuclphysb.2014.03.007}{\emph{Nucl. Phys. B}
  {\bfseries 882} (2014) 303--351},
  [\href{https://arxiv.org/abs/1401.7307}{{\ttfamily 1401.7307}}].

\bibitem{Braun:2014owa}
V.~M. Braun and A.~N. Manashov, \emph{{Conformal symmetry of the Lange-Neubert
  evolution equation}},
  \href{https://doi.org/10.1016/j.physletb.2014.02.051}{\emph{Phys. Lett. B}
  {\bfseries 731} (2014) 316--319},
  [\href{https://arxiv.org/abs/1402.5822}{{\ttfamily 1402.5822}}].

\bibitem{Korchemsky:1987wg}
G.~P. Korchemsky and A.~V. Radyushkin, \emph{{Renormalization of the Wilson
  Loops Beyond the Leading Order}},
  \href{https://doi.org/10.1016/0550-3213(87)90277-X}{\emph{Nucl. Phys. B}
  {\bfseries 283} (1987) 342--364}.

\bibitem{Eichten:1980mw}
E.~Eichten and F.~Feinberg, \emph{{Spin Dependent Forces in QCD}},
  \href{https://doi.org/10.1103/PhysRevD.23.2724}{\emph{Phys. Rev. D}
  {\bfseries 23} (1981) 2724}.

\bibitem{Matsuura:1988sm}
T.~Matsuura, S.~C. van~der Marck and W.~L. van Neerven, \emph{{The Calculation
  of the Second Order Soft and Virtual Contributions to the Drell-Yan
  Cross-Section}},
  \href{https://doi.org/10.1016/0550-3213(89)90620-2}{\emph{Nucl. Phys. B}
  {\bfseries 319} (1989) 570--622}.

\bibitem{Becher:2009kw}
T.~Becher and M.~Neubert, \emph{{Infrared singularities of QCD amplitudes with
  massive partons}},
  \href{https://doi.org/10.1103/PhysRevD.79.125004}{\emph{Phys. Rev. D}
  {\bfseries 79} (2009) 125004},
  [\href{https://arxiv.org/abs/0904.1021}{{\ttfamily 0904.1021}}].

\bibitem{Moch:2004pa}
S.~Moch, J.~A.~M. Vermaseren and A.~Vogt, \emph{{The Three loop splitting
  functions in QCD: The Nonsinglet case}},
  \href{https://doi.org/10.1016/j.nuclphysb.2004.03.030}{\emph{Nucl. Phys. B}
  {\bfseries 688} (2004) 101--134},
  [\href{https://arxiv.org/abs/hep-ph/0403192}{{\ttfamily hep-ph/0403192}}].

\bibitem{Henn:2019swt}
J.~M. Henn, G.~P. Korchemsky and B.~Mistlberger, \emph{{The full four-loop cusp
  anomalous dimension in $\mathcal{N}=4$ super Yang-Mills and QCD}},
  \href{https://doi.org/10.1007/JHEP04(2020)018}{\emph{JHEP} {\bfseries 04}
  (2020) 018}, [\href{https://arxiv.org/abs/1911.10174}{{\ttfamily
  1911.10174}}].

\bibitem{vonManteuffel:2020vjv}
A.~von Manteuffel, E.~Panzer and R.~M. Schabinger, \emph{{Cusp and collinear
  anomalous dimensions in four-loop QCD from form factors}},
  \href{https://doi.org/10.1103/PhysRevLett.124.162001}{\emph{Phys. Rev. Lett.}
  {\bfseries 124} (2020) 162001},
  [\href{https://arxiv.org/abs/2002.04617}{{\ttfamily 2002.04617}}].

\bibitem{Beneke:2009rj}
M.~Beneke, P.~Falgari and C.~Schwinn, \emph{{Soft radiation in heavy-particle
  pair production: All-order colour structure and two-loop anomalous
  dimension}},
  \href{https://doi.org/10.1016/j.nuclphysb.2009.11.004}{\emph{Nucl. Phys. B}
  {\bfseries 828} (2010) 69--101},
  [\href{https://arxiv.org/abs/0907.1443}{{\ttfamily 0907.1443}}].

\bibitem{Erdogan:2011yc}
O.~Erdo\u{g}an and G.~Sterman, \emph{{Gauge Theory Webs and Surfaces}},
  \href{https://doi.org/10.1103/PhysRevD.91.016003}{\emph{Phys. Rev. D}
  {\bfseries 91} (2015) 016003},
  [\href{https://arxiv.org/abs/1112.4564}{{\ttfamily 1112.4564}}].

\bibitem{Falcioni:2019nxk}
G.~Falcioni, E.~Gardi and C.~Milloy, \emph{{Relating amplitude and PDF
  factorisation through Wilson-line geometries}},
  \href{https://doi.org/10.1007/JHEP11(2019)100}{\emph{JHEP} {\bfseries 11}
  (2019) 100}, [\href{https://arxiv.org/abs/1909.00697}{{\ttfamily
  1909.00697}}].

\bibitem{Becher:2007ty}
T.~Becher, M.~Neubert and G.~Xu, \emph{{Dynamical Threshold Enhancement and
  Resummation in Drell-Yan Production}},
  \href{https://doi.org/10.1088/1126-6708/2008/07/030}{\emph{JHEP} {\bfseries
  07} (2008) 030}, [\href{https://arxiv.org/abs/0710.0680}{{\ttfamily
  0710.0680}}].

\bibitem{Beneke:2018rbh}
M.~Beneke, M.~Garny, R.~Szafron and J.~Wang, \emph{{Anomalous dimension of
  subleading-power $N$-jet operators. Part II}},
  \href{https://doi.org/10.1007/JHEP11(2018)112}{\emph{JHEP} {\bfseries 11}
  (2018) 112}, [\href{https://arxiv.org/abs/1808.04742}{{\ttfamily
  1808.04742}}].

\bibitem{Bell:2013tfa}
G.~Bell, T.~Feldmann, Y.-M. Wang and M.~W.~Y. Yip, \emph{{Light-Cone
  Distribution Amplitudes for Heavy-Quark Hadrons}},
  \href{https://doi.org/10.1007/JHEP11(2013)191}{\emph{JHEP} {\bfseries 11}
  (2013) 191}, [\href{https://arxiv.org/abs/1308.6114}{{\ttfamily 1308.6114}}].

\bibitem{Braun:2015pha}
V.~M. Braun, A.~N. Manashov and N.~Offen, \emph{{Evolution equation for the
  higher-twist B-meson distribution amplitude}},
  \href{https://doi.org/10.1103/PhysRevD.92.074044}{\emph{Phys. Rev. D}
  {\bfseries 92} (2015) 074044},
  [\href{https://arxiv.org/abs/1507.03445}{{\ttfamily 1507.03445}}].

\bibitem{Braun:2018fiz}
V.~M. Braun, Y.~Ji and A.~N. Manashov, \emph{{Integrability in heavy quark
  effective theory}},
  \href{https://doi.org/10.1007/JHEP06(2018)017}{\emph{JHEP} {\bfseries 06}
  (2018) 017}, [\href{https://arxiv.org/abs/1804.06289}{{\ttfamily
  1804.06289}}].

\bibitem{Braun:2019zhp}
V.~M. Braun, Y.~Ji and A.~N. Manashov, \emph{{Scale-dependence of the $B$-meson
  LCDA beyond leading order from conformal symmetry}},  in \emph{{14th
  International Symposium on Radiative Corrections}: {Application of Quantum
  Field Theory to Phenomenology}}, 12, 2019,
  \href{https://arxiv.org/abs/1912.03210}{{\ttfamily 1912.03210}},
  \href{https://doi.org/10.22323/1.375.0037}{DOI}.

\end{thebibliography}\endgroup

\end{document}